\documentclass[10pt]{article}
\usepackage{amets,amssymb,amsmath,amstext,lineno,color,xcolor,comment,grffile,array,multirow,makecell,nicefrac}
\usepackage[percent]{overpic}
\usepackage{hyperref,graphicx,multicol,pdfpages}
\hypersetup{colorlinks = true, allcolors = blue, allbordercolors = white}
\usepackage[round,authoryear]{natbib}
\usepackage{float,rotating,epic,tabularx}
\bibpunct[,]{(}{)}{;}{a}{}{,}
\begin{document}
\renewcommand{\baselinestretch}{1.0}\small\normalsize

\small
\noindent
The following is a preprint submitted to the journal Atmosphere-Ocean
for peer review on October 10, 2024 (\textcopyright\ 2025
under the
\href{https://creativecommons.org/licenses/by-nc-nd/4.0}{CC-BY-NC-ND
  4.0 license}).  A revised version was accepted on June 4, 2025.
The definitive version from the publisher employs the title
``A seasonal to decadal calibration of 1990--2100 eastern Canadian
  freshwater discharge simulations by observations, data models,
  and neural networks.''  All versions seek to address:
\vspace{0.3in}

\noindent
$\bullet$\ \ Calibration of the forcing (input), parameters (internal), and streamflow (output) of a hydrological model\\
$\bullet$\ \ Relatively direct calibration of CMIP-forced simulations to observations\\
$\bullet$\ \ Recognition of a familiar measurement modelling framework at each step\\

\begin{center}
\vspace{0.2in}
{\Large A seasonal to decadal calibration of 1990--2100 eastern\\
  Canadian freshwater discharge simulations by\\
  observations, data models, and neural networks\\}
\vspace{0.3in}

\small
\noindent
Richard E. Danielson$^{1,2}$, Minghong Zhang$^{2}$, Jo\"{e}l Chass\'{e}$^{1}$, and Will Perrie$^{2}$\\
$^{1}$Fisheries and Oceans Canada, Gulf Fisheries Centre, Moncton, New Brunswick;\\
$^{2}$Fisheries and Oceans Canada, Bedford Institute of Oceanography, Dartmouth, Nova Scotia, Canada\\
\end{center}
\vspace{0.1in}

\begin{center}
\vspace{0.3in}

\large
Abstract\\
\end{center}
\normalsize

  A configuration of the NCAR WRF-Hydro model was sought using well
  established data models to guide the initial hydrologic model setup,
  as well as a seasonal streamflow post-processing by neural networks.
  Discharge was simulated using an eastern Canadian river network at
  two-km resolution.  The river network was taken from a digital
  elevation model that was made to conform to observed catchment
  boundaries.  Perturbations of a subset of model parameters were
  examined with reference to streamflow from 25~gauged catchments
  during the 2019~warm season.  A data model defines the similarity of
  modelled streamflow to observations, and improvements were found in
  about half the individual catchments.  With reference to 183~gauged
  catchments (1990-2022), further improvements were obtained at
  monthly and annual scales by neural network post-processing that
  targets all catchments at once as well as individual catchments.

  This seasonal calibration was applied to uncoupled WRF-Hydro
  simulations for the 1990-2100 warming period.  Historic and future
  forcing were provided, respectively, by a European Centre for
  Medium-Range Weather Forecasting reanalysis (ERA5), and by a WRF
  atmospheric model downscaling of a set of Coupled Model
  Intercomparison Project (CMIP) models, where the latter were also
  seasonally calibrated.  Eastern Canadian freshwater discharge peaks
  at about 10$^5$~m$^3$~s$^{-1}$, and as previous studies have shown,
  there is a trend toward increasing low flows during the cold season
  and an earlier peak discharge in spring.  By design, neural networks
  yield more precise estimates by compensating for different
  hydrologic process representations.

\vspace{0.3in}
\noindent\rule{\textwidth}{0.5pt}

\newpage
\begin{multicols}{2}
\normalsize

\section{Introduction}

Anticipation of future changes in the water cycle depends in part on
hydrologic model simulations that are capable of reproducing
historical river discharge estimates.  \citet{Bush_Lemmen_etal_2019}
summarize climate projections for Canada, as well as ongoing changes
in the water cycle as global temperature rises.  Precipitation may
have already increased across northern Canada, and while less of this
is frozen, the total amount is expected to increase further through
2100, particularly in winter \citep{Bush_Lemmen_Zhang_etal_2019}.
Permafrost temperature is also increasing as the duration of land and
marine snow and ice cover decreases
\citep{Bush_Lemmen_Derksen_etal_2019}.  Cold season river discharge is
expected to continue increasing, with an earlier peak discharge in
spring \citep{Bush_Lemmen_Bonsal_etal_2019}.  Concurrently, surface
density stratification may be changing in Canadian waters
\citep{Bush_Lemmen_Greenan_etal_2019}.

The hydrologic component of the water cycle begins with precipitation
reaching the ground and ends with catchment loss, either by
evaporation or discharge to the ocean \citep{Beven_2019}.  Hydrologic
simulations are needed to assess the downstream impact of temperature
and precipitation changes on streamflow and ocean stratification.
Increasing and earlier peak river discharge trends are found by
\citet{Stadnyk_etal_2021} in Arctic simulations that cover 1981-2070
using the Hydrological Predictions for the Environment (HYPE) model of
the Swedish Meteorological and Hydrological Institute (SMHI).  As
hydrologic models are sensitive to bias in the forcing used for
calibration \citep{Berg_etal_2018}, atmospheric forcing for 1981-2010
is taken from the Hydrological Global Forcing Data (HydroGFD), which
in turn, is a calibration reference for 2011-2070 forcing by Earth
System models \citep{Taylor_etal_2012, IPCC_2013} of the Coupled Model
Intercomparison Project (CMIP).  Stadnyk et al.~also perform ocean
simulations for 2002-2009 using the Nucleus for European Modelling of
the Ocean (NEMO) model.  Their comparison of high-latitude forcing
from HYPE and from observed estimates of monthly-mean streamflow
\citep{Dai_etal_2009} points to a hydrologic contribution to ocean
stratification changes.

Components of Earth System models that provide operational forecasts
are known to perform well against observations
\citep{Schmidt_etal_2017}.  Like HYPE, the National Center for
Atmospheric Research (NCAR) Weather Research and Forecasting (WRF)
hydrologic model (WRF-Hydro; \citealt{Gochis_etal_2021}) is employed
in operations.  The natural representations that WRF-Hydro employs
include column land surface (NOAH-MP), subsurface flow, and overland
flow components that drive the main discharge to the ocean via a river
and lake network \citep{Gochis_etal_2021}.  Streamflow routing follows
a river network that the WRF-Hydro preprocessor interpolates from a
digital elevation model (DEM).  The subkilometer resolution of most
DEMs may be appropriate in operations, but \citet{Eilander_etal_2021}
provides an upscaled DEM that preserves river structure at lower
resolution.  Moreover, \citet{Lehner_etal_2008} provide watershed
boundaries to preserve an interpolated catchment area upstream of
hydrologic stations \citep{deRham_etal_2020,
  Pellerin_NzokouTanekou_2020}.  With a river network that reproduces
the natural network as closely as possible, both in terms of station
location and upstream area, it is easier to associate WRF-Hydro
simulations with observations.

A calibration to observations can be seen as {\it anchoring} a
hydrologic model when both its internal processes, and the variables
it shares with adjacent Earth system models, more closely mimic nature
\citep{Schmidt_etal_2017, Stadnyk_etal_2020}.  Regarding upstream
atmospheric forcing, its calibration is expected to depend on the
individual CMIP model \citep{Berg_etal_2018}, but also should allow
for the synoptic evolution of a freely evolving climate model, which
is different than in observations \citep{Maraun_2016}.  Not only does
multivariate bias correction address such upstream issues, but
notably, it also preserves decadal trends \citep{Cannon_2018},
although the temporal sequence of CMIP forcing is adjusted slightly to
match the probability distributions of observations.  We return to
this issue in Section~3.d, where we seek to associate simulated and
observed streamflow without matching synoptic sequences.

Regarding internal parameters of a hydrologic model (and Earth system
models generally), the identification of unique values for an
increasing number of parameters is another known challenge
\citep{Schmidt_etal_2017, Beven_2021}.  If computer resources are well
allocated \citep{Wang_etal_2019}, then automatic calibration offers a
partial solution \citep{Doherty_2015}.  A comprehensive calibration of
parameters in HYPE is given by \citet{Stadnyk_etal_2020}, who develop
multiyear simulations of discharge to Hudson Bay, with representations
of lake storage and frozen soil.  Examples of automatic calibration of
WRF-Hydro include \citet{Wang_etal_2019} and
\citet{RafieeiNasab_etal_2020}, who rely on objective functions
similar to those of Stadnyk et al.~to identify model parameters.

\end{multicols}
\begin{table}
  \begin{center}
  {\begin{tabular}{cccccc}
  \hline
  Step & \begin{tabular}{@{}c@{}}Process\\Model\end{tabular} & Representation & Reference  & \begin{tabular}{@{}c@{}}Data\\Model\end{tabular} & Calibration \\
  \hline
    1  & Topographic & Watershed boundaries    & HydroSHEDS & Truth  & Section 3a \\
    1  & Atmospheric & Downscaled CMIP forcing & ERA5       & Truth  & Section 3b \\
    3  & Hydrologic  & WRF-Hydro parameters    & HyDAT      & Target & PEST   \\
    4  & Hydrologic  & WRF-Hydro streamflow    & HyDAT      & Target & Neural network \\
  \hline
  \end{tabular}}
  \caption{Summary of calibration steps applied in this study and the type
    of representation involved, along with the choice of reference,
    data model, and method of calibration (see text for definitions).}
  \end{center}
  \label{tab01}
\end{table}
\begin{multicols}{2}

Regarding downstream ocean forcing, calibration of the output of a
hydrologic model can help to address river regulation and streamflow
buffering by many small lakes in eastern Canada \citep{Dery_etal_2018,
  Stadnyk_etal_2020}.  If observations are available, then advances in
neural network training \citep{Rumelhart_etal_1986, Innes_etal_2018}
enable automatic calibration of generic model structures, which can be
used to relate simulated and observed streamflow at gauged stations
furthest downstream, and at eastern Canadian ocean outlets
\citep{Dai_Trenberth_2002}.  \citet{Stiles_etal_2014} demonstrates
that a composition of neural networks can be used to target different
aspects of a relationship.  However, because they offer a
parameterized nonlinear relationship between WRF-Hydro and HyDAT
streamflow, it is not obvious that decadal trends are preserved
\citep{Maraun_2016, Cannon_2018}.

Hydrologic models are expected to require calibration following any
change in configuration, including going from an operational
resolution of one km or less to climate simulations at coarser
resolution.  Within an Earth system modelling framework, each
calibration step upstream, internal, and downstream of WRF-Hydro is
important.  The steps taken in this study are shown in
Table~1, beginning with enforcing watershed boundaries and
with a seasonal calibration of atmospheric forcing.  This is followed
by hydrologic calibration using an automatic parameter estimation tool
called PEST \citep{Doherty_2015, Wang_etal_2019}.  Calibration of
WRF-Hydro streamflow is then explored using neural networks
\citep{Stiles_etal_2014, Innes_etal_2018}.  As ocean forcing is
emphasized below, we focus increasingly on the hydrologic steps, using
observed discharge from eastern Canadian rivers as our reference.
Methods and results are provided for each calibration step in
Sections~3 and~4, respectively.  A discussion is given in Section~5
and conclusions are given in Section~6.  The next section describes
the observed and simulated data.

\section{Data}

\begin{figure}
\begin{center}
  \includegraphics[width=0.73\textwidth]{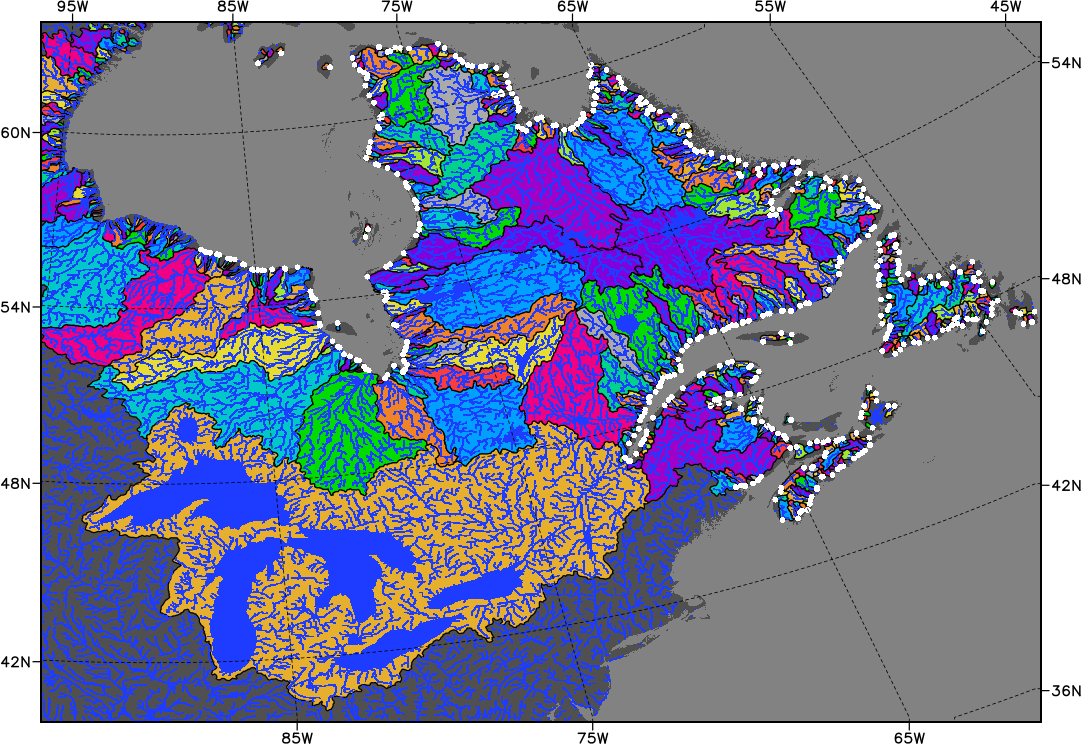}
  \caption{Native WRF-Hydro projection (Lambert conformal conic) of
    the eastern Canadian 50-km (54$\times$62) land surface model and 2-km
    (1350$\times$1550) river routing grids, with a river and lake network
    (dark blue) that discharges from watersheds (colours) to the ocean
    at 477~outlets between the Severn (Ontario) and St.~Croix (New
    Brunswick/Maine) rivers (white dots).}
  \label{fig01}
\end{center}
\end{figure}

Hydrologic processes are represented by WRF-Hydro on a grid projection
at two resolutions (Fig.~\ref{fig01}).  We obtain unadjusted
atmospheric and topographic data from a western North Atlantic
downscaling of multi-year CMIP simulations \citep{Zhang_etal_2019b}
and a global upscaled DEM \citep{Eilander_etal_2021} at horizontal
grid resolutions of 30~km and 1~km, respectively.  Slightly lower
resolutions of 50~km and 2~km are selected for the WRF-Hydro land
surface model (54 by 62) and river routing (1350 by 1550) grids
covering eastern Canada.

\subsection{Topographic data}

Surface elevation from satellites requires adjustment before a river
network can be derived that matches observed paths and flow directions
\citep{Lehner_etal_2008, Yamazaki_etal_2019}.  The WRF-Hydro
preprocessor performs bilinear interpolation of a DEM to the routing
grid, but the resulting river network and watershed boundaries are
sensitive to the input DEM resolution.  Using the hydrologically
adjusted DEM of \citet{Yamazaki_etal_2019}, \citet{Eilander_etal_2021}
develop an iterative approach to representing river structure, length,
and slope at lower resolution.  Reliance on interpolation is reduced
as this DEM is at slightly higher resolution than the WRF-Hydro
routing grid.  To ensure that catchment areas upstream of hydrologic
stations are also preserved, watershed boundaries in the 15-arcsec
HydroSHEDS DEM \citep{Lehner_etal_2008} are employed as a reference.

\end{multicols}
\newpage
\begin{multicols}{2}

\subsection{Atmospheric data}

A useful reference for historical CMIP simulations is a reanalysis of
observed atmospheric forcing \citep{Stadnyk_etal_2021}.  We employ the
fifth European Centre for Medium-Range Weather Forecasting Reanalysis
(ERA5; \citealt{Hersbach_etal_2020}), which is based on a spectral
model (137~vertical levels and an effective horizontal resolution of
31~km) with data assimilation over successive 12-h periods, and a
reduction in systematic differences between the model and observations
\citep{Dee_2005}.  This global atmospheric reference is available
hourly on a 0.25$^\circ$~grid, and we bilinearly interpolate
three-hourly surface forcing for 1990-2022 to the 50-km WRF-Hydro grid
(Fig.~\ref{fig01}).

Our representations of historical and future forcing are a combination
of low resolution CMIP simulations \citep{Bush_Lemmen_Flato_etal_2019}
and enhanced (30-km) resolution by a Weather Research and Forecasting
(WRF) model downscaling \citep{Zhang_etal_2019b}.  A standard
hydroclimatic assessment may employ~15 representations or more, but we
employ four, where WRF is driven at lateral and lower boundaries by a
selected Earth System simulation that, in turn, is driven by a shared
socioeconomic pathway (SSP2-4.5 or SSP5-8.5) scenario
\citep{ONeill_etal_2017}.  We employ three simulations from the CMIP5
experiment \citep{Taylor_etal_2012} and one from CMIP6
\citep{Eyring_etal_2016}.  Respectively, these are an NCAR Community
Climate System Model (CCSM-4 SSP5-8.5) simulation
\citep{Meehl_etal_2012}, two Met Office Hadley Centre Global
Environmental Model (HadGEM2 SSP2-4.5 and SSP5-8.5) simulations
\citep{Collins_etal_2011}, and a Max Planck Institute for Meteorology
Earth System Model (MPI-ESM1.2-LR SSP5-8.5) simulation
\citep{Mauritsen_etal_2019}.  Each six-hourly representation for
1990-2100 is also bilinearly interpolated to the 50-km grid.

\subsection{Hydrologic data}

Instrumental daily streamflow serves as a useful reference for
calibrating WRF-Hydro parameters and output, although from day to day,
streamflow itself is not measured at most eastern Canadian hydrologic
stations.  Instead, river stage (i.e., water level) is monitored and
daily streamflow is obtained from stage-discharge relationships that
depend in part on open-water or ice-affected conditions
\citep{deRham_etal_2020}.  These streamflow observations are taken
from HyDAT, the National Water Data Archive of the \citet{Hydat_2023}.
For WRF-Hydro parameter calibration (Section~3.c), we focus on
Reference Hydrometric Basin Network (RHBN) stations
\citep{Pellerin_NzokouTanekou_2020}, which are a HyDAT subset with
long records ($>$20 years) and little or no impact from upstream dams
and reservoirs.

\section{Methods}

We begin with the conventional assumption that calibrations based on
historical data can be applied well into the future
\citep{Maraun_2016}.  Two initial calibration steps are done in
parallel (Table~1) that seek to reproduce a natural river
network and perform a seasonal atmospheric calibration.  We then
perform hydrologic calibration, first using a parameter estimation
tool called PEST \citep{Doherty_2015, Wang_etal_2019}, and next by
discharge post-processing using neural networks
\citep{Stiles_etal_2014, Innes_etal_2018}.  These two steps are
similar in that discharge from large catchments involves nonlinear
contributions by many concurrent processes, and both PEST and neural
networks accommodate process nonlinearity.  Moreover, both calibration
steps are performed with reference to daily streamflow
\citep{Pellerin_NzokouTanekou_2020} and employ the same data model.
This data model is manifest in the PEST objective function and the
neural network loss function, where minima in the difference between
predicted and observed streamflow are sought.  Data models are defined
in the Appendix, with the calibrated reference and uncalibrated
representation (e.g., HyDAT observed and WRF-Hydro predicted
streamflow) denoted by $C$ and $U$, respectively.

\subsection{Topographic calibration}

To compare streamflow at HyDAT stations and on the WRF-Hydro routing
grid, we ensure that upstream catchment areas are comparable.  Thus,
watershed boundaries are located in the 15-arcsec HydroSHEDS DEM
\citep{Lehner_etal_2008} and the same boundaries are imposed in
WRF-Hydro by raising DEM values by O[100~m].  This is done before
interpolation and upstream of selected inland stations in Section~4.c,
and after interpolation and upstream of all outlets to the ocean in
Section~4.d.  Either inland or coastal adjustment is employed (not
both) to make the WRF-Hydro river network and watersheds similar to
HydroSHEDS.  For this calibration, it suffices to treat the HydroSHEDS
boundaries ($C$) as if they were fixed and natural.  The truth model
is employed, but we only confirm visually that the distance between
$C$ and $U$ is smaller when the HydroSHEDS boundaries are imposed
(Section~4.a).

\subsection{Atmospheric calibration}

A seasonal calibration using ERA5 as a reference is applied to
atmospheric forcing after interpolation to the WRF-Hydro 50-km land
surface grid (Section~2.b).  In addition to CMIP model downscaling
using the same atmospheric model \citep{Zhang_etal_2019b}, this
calibration helps to address systematic differences among CMIP models
\citep{Bush_Lemmen_Flato_etal_2019}.  Averages are taken over land in
Fig.~\ref{fig01} and for each day of the year between 1990 and 2004,
and the resulting annual cycle is approximated by a sinusoid
\citep{Jacquelin_2014}.  These smooth annual cycles define a daily
linear adjustment to ERA5 that is applied to all years (1990-2100).
Calibration is multiplicative for wind speed and precipitation,
additive for other variables, and some forcing variables are omitted
if a more sophisticated calibration is required (i.e., for shortwave
and longwave radiation), or differences with ERA5 are small.  Again
for this calibration, it suffices to treat the smooth annual cycle of
ERA5 ($C$) as fixed and natural.  The truth model is also employed and
we confirm visually that $C - U$ is smaller after calibration
(Section~4.b).

\subsection{Hydrologic model calibration}

The WRF-Hydro model is a parallelized, distributed process model
\citep{Gochis_etal_2021}, whose core components (and our specific
settings) are a column land surface model (four-layer Noah-MP),
overland flow (steepest descent), shallow subsurface flow (saturated
flow), groundwater baseflow (exponential bucket), channel routing
(diffusive wave), and reservoir routing (level-pool).  Simulations
employ land surface and river routing timesteps of 1~h and 30~sec,
respectively.  The automatic parameter estimation tool PEST
\citep{Doherty_2015} is employed to tune WRF-HYDRO parameters to yield
a similarity to observations.  For the conterminous U.S.,
\citet{RafieeiNasab_etal_2020} provide the range and sensitivity of a
number of WRF-Hydro parameters, of which nine are selected for
calibration (Table~2).

This calibration ignores CMIP data, and instead estimates parameters
using the coordinated synoptic evolution of ERA5 forcing and HyDAT
streamflow.  A three-year WRF-Hydro spinup is performed using
annual-mean ERA5 forcing, and a baseline simulation is performed from
October 1, 2018 to October 31, 2019 using default parameters.
Following \citet{Wang_etal_2019}, over 100~WRF-Hydro simulations are
then launched during parameter estimation.  Each begins with a restart
from the baseline simulation on March~31, 2019.  The PEST objective
function ($\Phi_{P}$) is evaluated using model and observed streamflow
differences between May and October.  Apart from WRF-Hydro itself, the
only process (i.e., ``expert'') knowledge given to PEST is the
acceptable range in each free parameter
\citep{RafieeiNasab_etal_2020}.  In other words, we employ PEST's
``estimation mode'' and omit prior information and additional process
knowledge (i.e., Tikhonov regularization).  The objective function
minimization employs singular value decomposition (SVD) for numerical
stability \citep{Doherty_2015}.

\end{multicols}
\begin{table}
  \begin{center}
  {\begin{tabular}[l]{lcrc}
  \hline
  Parameter    & Description                                   &      Range           & PEST Value \\
  \hline
  BEXP         & Coefficient of pore size distribution         & $\times$ [0.4,  1.9] & {\bf 0.47} \\
  DKSAT        & Saturated hydraulic conductivity (m s$^{-1}$) & $\times$ [0.2, 10.0] & {\bf 0.21} \\
  MFSNO        & Snow depletion melt                           &          [0.5,  3.0] & {\bf 0.50} \\
  MP           & Slope of Ball-Berry conductance               & $\times$ [0.6,  1.4] & {\bf 1.40} \\
  OVROUGHRTFAC & Manning's roughness for overland flow         &          [0.5,  1.5] & {\bf 1.50} \\
  REFKDT       & Infiltration/surface runoff partitioning      &          [0.1,  4.0] &      0.68  \\
  RETDEPRTFAC  & Maximum retention depth                       &          [0.1, 10.0] & {\bf 0.23} \\
  SLOPE        & Bottom drainage boundary                      &          [0.0,  1.0] &      0.27  \\
  SMCMAX       & Saturated soil moisture content               & $\times$ [0.8,  1.2] &      1.02  \\
  \hline
  \end{tabular}}
  \caption{Nine WRF-Hydro calibration parameters, as in
    \citet{RafieeiNasab_etal_2020}, with corresponding descriptions,
    ranges, and PEST values (bold values are close to a range limit;
    see Section~4.c).  All parameter values are spatially constant,
    but ranges that begin with $\times$ multiply a spatially varying
    default field.}
  \end{center}
\label{tab02}
\end{table}
\begin{multicols}{2}

The guided search for WRF-HYDRO parameters (Table~2) leads
to a greater similarity to streamflow observations, which is a well
established definition of improvement.  Although PEST and neural
networks are different methods of calibration, when deriving their
objective or loss function, they share the same target data model and
make the same {\it equivalence assumption} between $C$ and $U$ (see
Appendix).  For PEST, the basic function that guides the parameter
search is
\begin{equation}
  \Phi_{P} = [C – U]^T Q^{-1} [C – U],
  \label{optp}
\end{equation}
where $C$ and $U = Xp$ are streamflow ($X$ is a linear representation
of the WRF-Hydro model and $p$ contains the parameter values), and $Q$
corresponds to a HyDAT and WRF-Hydro error covariance matrix.  We
consider minimization of $\Phi_{P}$ to be complete after five
iterations, each of which usually employs more than 20~WRF-Hydro
simulations \citep{Doherty_2015}.

\subsection{Hydrologic data calibration}

The specific configuration of WRF-Hydro that is employed in this study
may limit the physical processes that are simulated, and we would not
expect a subsequent calibration to compensate for missing processes.
Notable challenges in eastern Canada include the prevalence of small
lakes, frozen soil, and river regulation \citep{Dery_etal_2018,
  Stadnyk_etal_2020}.  Insofar as processes are lacking in our
simulations, however, we also expect these to have a systematic
nonlinear impact on the processes that are both a)~captured by HyDAT
observations, and b)~simulated by WRF-Hydro but without being
impacted.  It is by way of a parameterized nonlinear relationship
between WRF-Hydro and HyDAT streamflow that we propose a subsequent
calibration of WRF-Hydro output.

Neural networks are adaptive and convenient for identifying a nonlinear
relationship when some of the physical processes involved are not
apparent.  In addition to process differences, CMIP model forcing is
freely evolving, so WRF-Hydro and HyDAT streamflow are asynchronous on
synoptic scales.  \citet{Stiles_etal_2014} demonstrates that a
composition of neural networks can capture different aspects of a
relationship, so two neural networks are employed.  The first (NN232
hereafter) seeks to ``buffer'' streamflow by reducing strong peaks in
both seasonal and daily flows.  Thus, we design its input and output
layers to have two nodes for daily streamflow and a 12-day centered
average.  Between these layers is a three-node hidden layer with batch
normalization, a dense connection between all layers, and an ELU
activation function at each connection \citep{Clevert_etal_2015}.

The second neural network (NN343 hereafter) provides a nonlinear
parameterization of a)~missing processes like river regulation and
b)~the freely evolving synoptic-scale forcing of each CMIP model.  We
design its input and output layers to have three nodes for daily,
monthly, and annual-mean streamflow, with dense connections to a
four-node hidden layer with batch normalization.  Of course, (b) does
not pertain to ERA5 forcing, since WRF-Hydro and HyDAT streamflow
would be synchronized.  However, a synoptic pairing of daily data is
not possible when training separately for each CMIP model, and in that
case, we allow for an asynchronous synoptic evolution.  Following
\citet{Freilich_Challenor_1994}, ranked daily flows in WRF-Hydro and
HyDAT are matched (but the order of the monthly and annual-mean input
and output data are unchanged).  The result is a seasonally calibrated
streamflow whose temporal sequence is unaltered
(cf.~\citealt{Maraun_2016, Cannon_2018}).

\newpage
\end{multicols}
\begin{figure}
\begin{center}
  \includegraphics[width=0.49\textwidth]{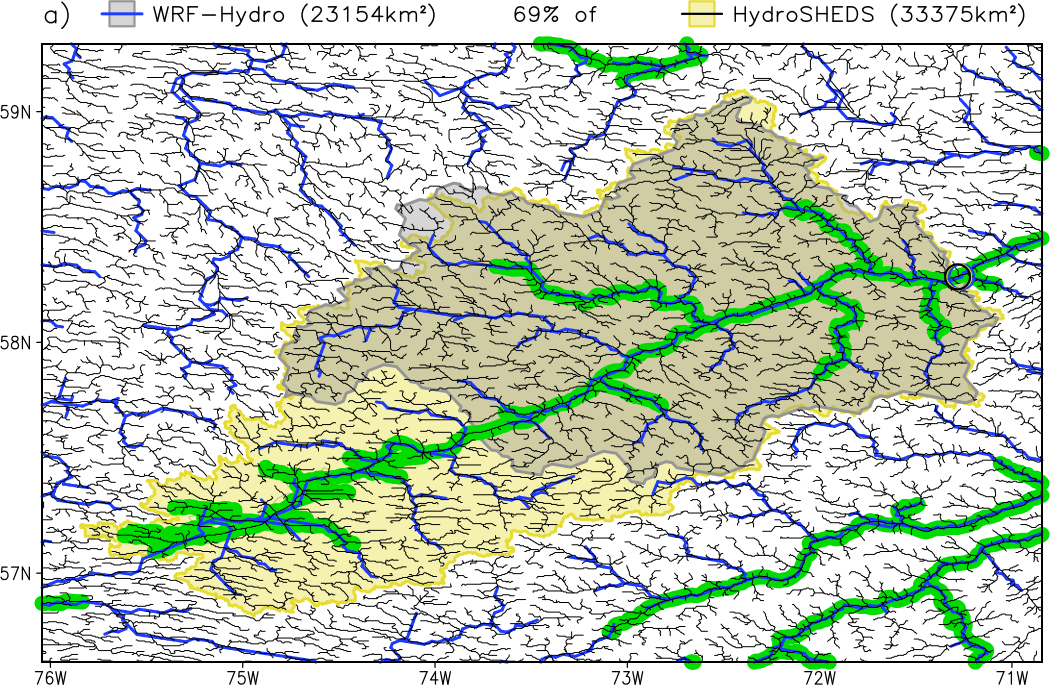}
  \hspace{0pt}
  \includegraphics[width=0.49\textwidth]{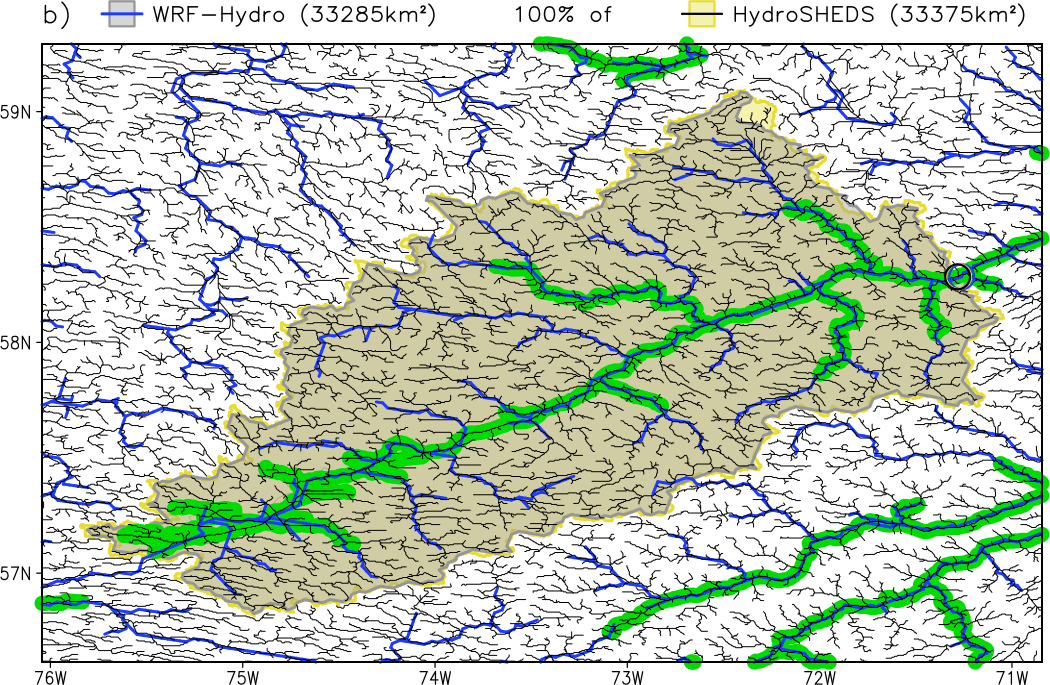}
  \caption{Catchment area upstream of hydrologic station 03JB004
    (Rivi{\`e}re Aux Feuilles, upstream of Dufreboy Brook) following
    an interpolation of the 30-arcsec DEM of
    \citet{Eilander_etal_2021} to the 2-km WRF-Hydro grid a)~before
    and b)~after raising the height of the input DEM by 100~m at the
    15-arcsec HydroSHEDS catchment boundary \citep{Lehner_etal_2008}.
    Shown are the station (circle), WRF-Hydro rivers (blue) and
    catchments (grey), HydroSHEDS rivers (black) and catchments
    (yellow), and rivers wider than about 30~m from Landsat images
    (green; \citealt{Altenau_etal_2021}).}
  \label{fig02}
\end{center}
\end{figure}
\begin{multicols}{2}

The number of parameters sought for NN232 (23) and NN343 (39) is more
than for WRF-Hydro (9), and as in (\ref{optp}), guided searches lead
to a greater similarity to observations.  A similar loss function
($\Phi_{N}$) is also used for training:
\begin{equation}
  \Phi_{N} = [C – U]^T [C – U].
  \label{optn}
\end{equation}
Parameter solutions are obtained using Flux.jl \citep{Innes_etal_2018}
and we consider minimization complete after 1000 iterations.  Although
neural networks provide two or three outputs (e.g., daily, monthly,
and annual-mean), only daily streamflow is retained.  Finally, because
neural networks are trained at inland gauged stations, we also apply
them to WRF-Hydro discharge at the coast, where they provide a
nonlinear calibration (cf.~\citealt{Dai_Trenberth_2002}).

\subsection{Hydrologic performance}

One composite measure of performance is employed, following
\citet{Moriasi_etal_2015}, who define a satisfactory comparison at a
gauged station as having a difference (DIFF) between WRF-Hydro and
HyDAT of 15\% or less and Nash-Sutcliffe efficiency (NSE) of 0.5 or
greater.  The individual expressions of accuracy and skill are DIFF =
$\Sigma|U-C|/\Sigma|C|$, NSE = $1-\Sigma(U-C)^2/\Sigma(C-\bar{C})^2$,
root-mean-square difference (RMSD) = $(\Sigma(U-C)^2/n)^{0.5}$, bias
(BIAS) = $\Sigma(U-C)/n$, and Pearson correlation (COR) =
$\Sigma(U-\bar{U})(C-\bar{C}) / (\Sigma(C-\bar{C})^2
\Sigma(U-\bar{U})^2)^{0.5}$.  As in (\ref{optp}) and (\ref{optn}), all
but Pearson correlation seem to take $C$ and $U$ to be equivalent.

\section{Results}

\subsection{Topographic calibration}

Interpolation of the 30-arcsec hydrologically conditioned DEM of
\citet{Eilander_etal_2021} to the 2-km WRF-Hydro routing grid did not
always ensure that catchment area upstream of a hydrologic station or
an ocean outlet was fully captured.  An example of topographic
processing for hydrologic station 03JB004 (Rivi{\`e}re Aux Feuilles,
upstream of Dufreboy Brook) is shown in Fig.~\ref{fig02}a, but
catchments at other stations also differed (not shown).  The 30-arcsec
and 2-km resolutions are close, so much of the effort to preserve
river structure from higher resolution \citep{Eilander_etal_2021} was
retained on the WRF-Hydro grid.  This is apparent by the overlap among
three different river networks.  Imposing HydroSHEDS catchment
boundaries (Fig.~\ref{fig02}b) allowed an area occupied by Qasigialik
(Lake Minto) to drain eastward to Ungava Bay instead of westward to
Hudson Bay.  HydroSHEDS errors notwithstanding, we ensured that runoff
would be routed to the appropriate ocean outlet (Fig.~\ref{fig01}) to
support WRF-Hydro and HyDAT streamflow associations.

\newpage
\end{multicols}
\begin{table}
  \begin{center}
  {\begin{tabular}[l]{lccc}
  \hline
  WRF-Hydro forcing               & CCSM-4 SSP5-8.5 & HadGEM2 SSP2-4.5/5-8.5 & MPI-ESM1.2-LR SSP5-8.5 \\
  \hline
  Precipitation (mm hr$^{-1}$)     & $\times$ [1.0, 1.2]     & $\times$ [0.9, 1.2]             &                      \\
  Temperature ($^\circ$C)          &        - [0.3, 2.7]     &        + [2.1, 2.9]             &                      \\
  Specific Humidity (g kg$^{-1}$)  &        - [0.2, 0.6]     &                                 &                      \\
  Wind Speed (m s$^{-1}$)          & $\times$ [0.9, 1.0]     & $\times$ [0.8, 0.9]             &  $\times$ [1.0, 1.1] \\
  Surface Pressure (hPa)           &        - [0.8, 3.1]     &                                 &                      \\
  \hline
  \end{tabular}}
  \caption{Range of daily linear adjustments applied to downscaled CMIP
    simulations, where +/- and $\times$ refer to additive and multiplicative
    adjustments, respectively.  Shortwave and longwave radiation (not
    adjusted) are the remaining WRF-Hydro forcing variables.}
  \end{center}
\label{tab03}
\end{table}

\begin{figure}
\begin{center}
  \includegraphics[width=1.0\textwidth]{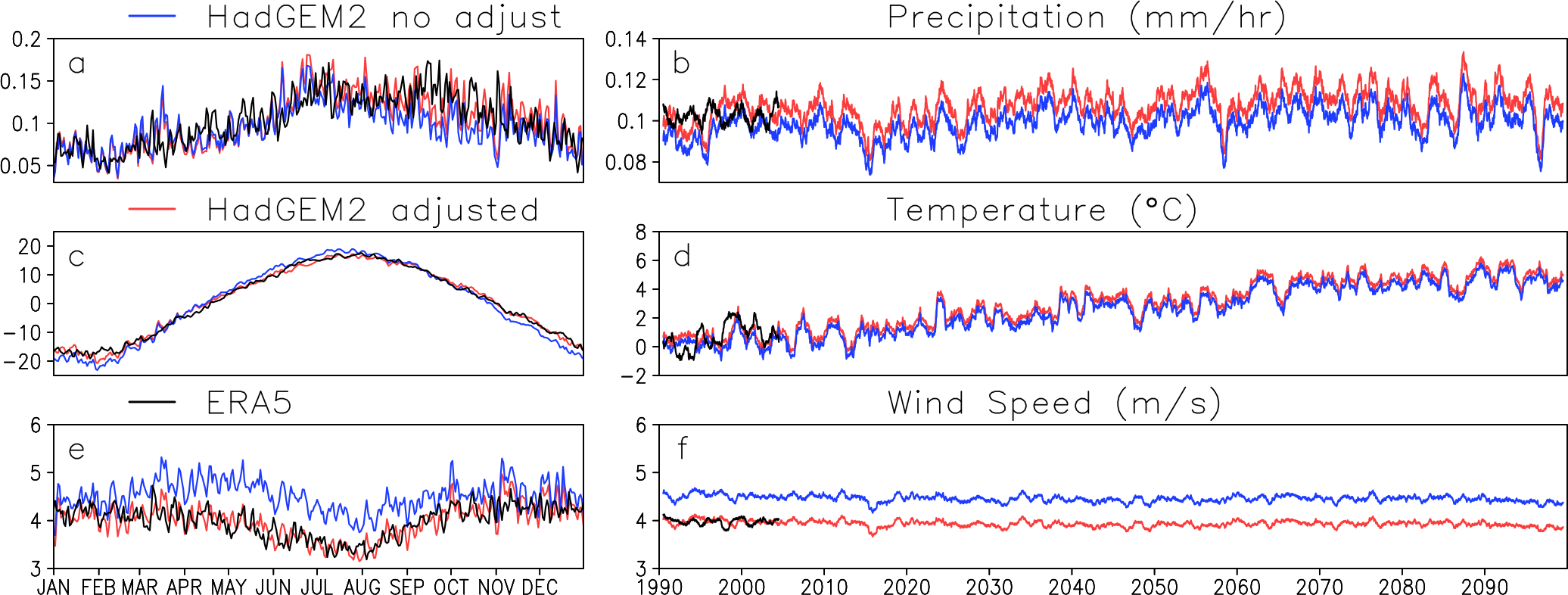}
  \caption{Linear adjustment of HadGEM2 to ERA5 (left panels) based on
    averages over land in Fig.~\ref{fig01} and for each day of the
    year between 1990 and 2004, and (right panels) applied to
    SSP2-4.5 annual averages for 1990-2100.  Shown are
    a,b)~precipitation (mm~hr$^{-1}$), c,d)~temperature ($^\circ$C),
    and e,f)~wind speed (m~s$^{-1}$). Note that the same historical
    adjustments are applied to HadGEM2 SSP5-8.5.}
  \label{fig03}
\end{center}
\end{figure}
\begin{multicols}{2}

\subsection{Atmospheric calibration}

Table~3 lists the range in additive and multiplicative
adjustments that were applied to the downscaled atmospheric forcing of
WRF-Hydro.  Five, three, and one WRF-Hydro variables were adjusted for
CCSM-4, HadGEM2, and MPI-ESM1.2-LR, respectively, with no adjustment
of shortwave and longwave radiation.  Adjustments for HadGEM2 are
shown in Fig.~\ref{fig03}, which reveals a deficit in precipitation
during September and October, and a warm-season surplus in wind speed
(blue lines in Fig.~\ref{fig03}a,e), whose adjustments (red lines) are
reflected in annual averages (Fig.~\ref{fig03}b,f).  Relative to ERA5,
CCSM-4 precipitation was also deficient in fall (not shown).  Although
we ignored ERA5 errors (e.g., any wind speed reference might be biased
low in storms), as expected, linear adjustments preserved the timing
of synoptic forcing (i.e., at 1-10~days) and simulated trends in
eastern Canada \citep{Maraun_2016, Bush_Lemmen_etal_2019}.

\subsection{Hydrologic model calibration}

\begin{figure}
\begin{center}
  \includegraphics[width=0.76\textwidth]{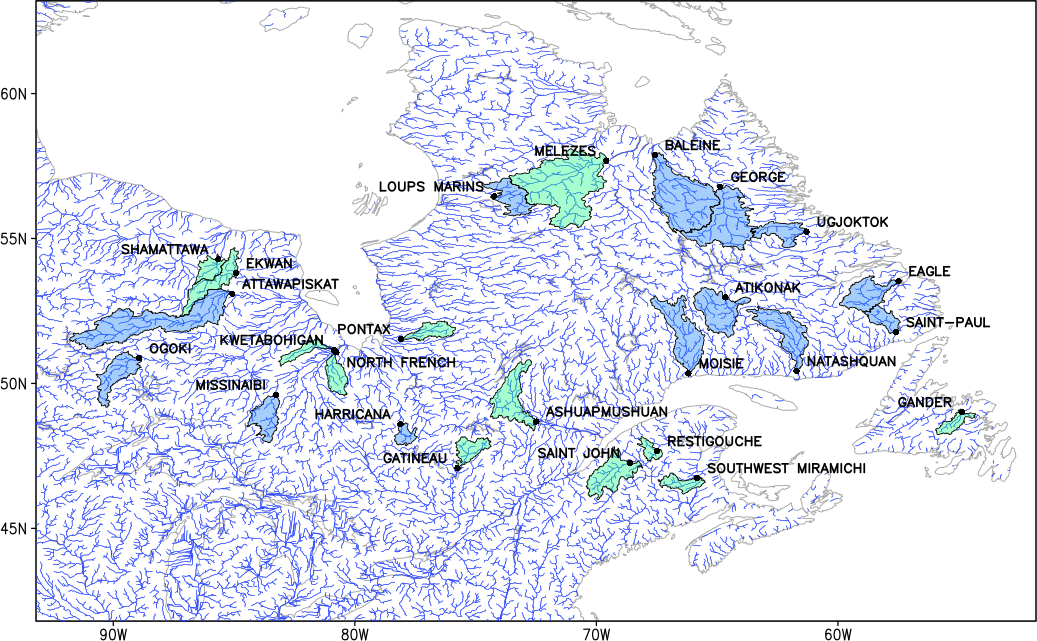}
  \caption{Catchments (shaded) upstream of 25~Reference Hydrometric
    Basin Network stations (black dots), where 3699~daily observations
    are available between May~31 and October~30, 2019.  Catchment
    shading denotes an increase (blue) or decrease (green) in
    WRF-Hydro similarity to observations (i.e., Nash-Sutcliffe
    efficiency) when going from default to PEST values for nine
    WRF-Hydro parameters (Table~2).  Each station provides
    153~observations, except for 03BF001/Pontax (63) and
    03KC004/M{\'e}l{\`e}zes (117).  Also shown is the 2-km WRF-Hydro
    river network with lakes omitted (blue lines).}
  \label{fig04}
\end{center}
\end{figure}

We sought a representation of {\it natural flow} through our river
network in an automatic calibration of WRF-Hydro parameters.  As the
Reference Hydrometric Basin Network (RHBN) stations are HyDAT stations
with little or no impact from upstream dams and reservoirs
\citep{Pellerin_NzokouTanekou_2020}, we selected 25~RHBN stations with
the largest upstream catchments (Fig.~\ref{fig04}), where
3699~streamflow observations were taken between the end of May and
October 2019.  The WRF-Hydro watershed boundaries were adjusted to
match HydroSHEDS (Section~3.a), lakes were omitted, and ERA5
atmospheric forcing was employed at 6-h intervals.

\newpage
\end{multicols}
\begin{multicols}{2}

Nine WRF-Hydro parameters (Table~2) were calibrated using
PEST, and although all are allowed to vary spatially
\citep{Gochis_etal_2021}, we employed (or multiplied by) constant
values across the eastern Canadian domain.  This calibration was
challenging because other WRF-Hydro parameters were fixed, including
Manning's roughness coefficients for rivers
(cf.~\citealt{Wang_etal_2019}).  We initially assumed, {\it but did
  not confirm}, that a natural-flow representation could be identified
for the entire eastern Canadian domain, as many parameters given by
PEST were close to their limit in range of application (bold values in
Table~2).  Moreover, in going from default to PEST parameter
values, we obtained greater similarity to observations (i.e.,
Nash-Sutcliffe efficiency) at only 13~of the 25~RHBN stations
(Fig.~\ref{fig04}).

A comparison of streamflow at individual stations revealed that the
HyDAT/RHBN observations varied more smoothly than WRF-Hydro
predictions, which captured strong peaks in both seasonal and daily
flows using default and PEST parameters.  Figure~\ref{fig05} is an
example of streamflow at the Attawapiskat River station
(Figure~\ref{fig04}), where default parameters overestimate the spring
freshet in May and PEST parameters overestimate two peaks in July.  A
simulation of strong daily flows seems consistent with our PEST values
of low soil infiltration (REFKDT) and retention (RETDEPRTFAC), and
high surface roughness (OVROUGHRTFAC; Table~2), but also,
RHBN stations capture streamflow buffering by small lakes upstream
\citep{Stadnyk_etal_2020}, whereas no lakes were included in our PEST
simulations.

\subsection{Hydrologic data calibration}

\begin{figure}
\begin{center}
  \includegraphics[width=0.75\textwidth]{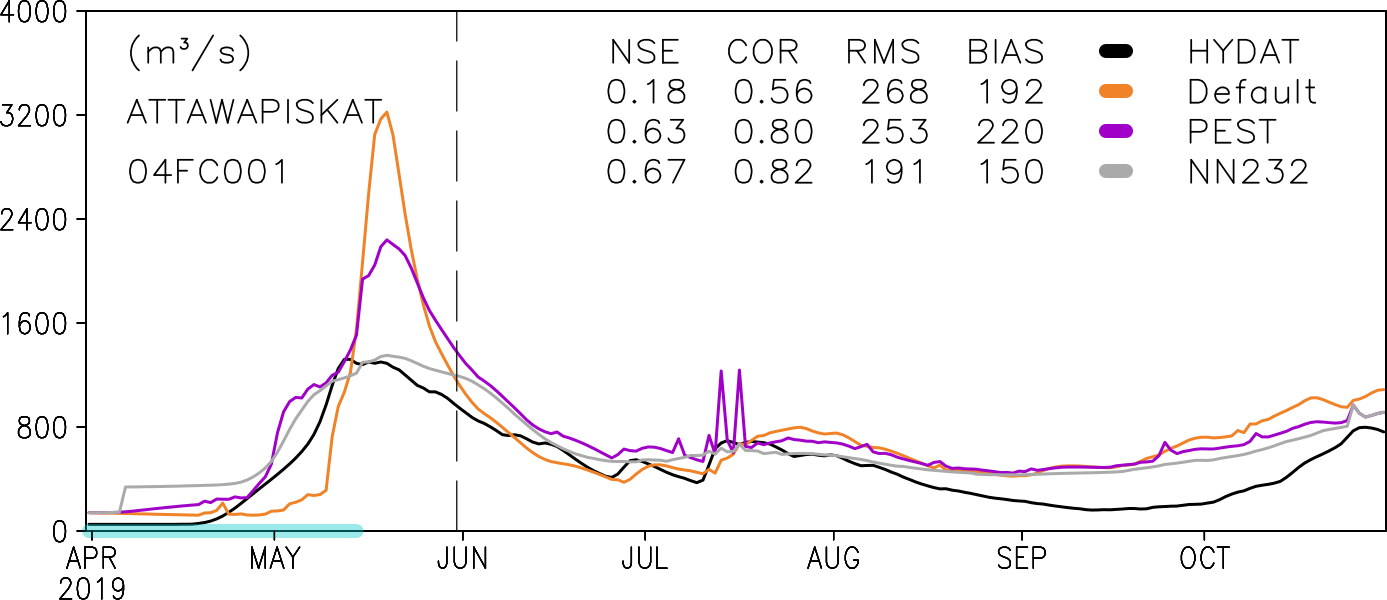}
  \caption{Daily discharge (m$^3$~s$^{-1}$) during 2019 at Attawapiskat
    River below Muketei River (04FC001), as given by HyDAT (black) and
    WRF-Hydro, where the latter employs default (orange) and PEST
    (purple) parameters, and after PEST dischange is post-processed
    using a neural network (NN232; grey).  Included are values of
    Nash-Sutcliffe efficiency (NSE), Pearson correlation (COR),
    root-mean-square difference (RMS), and bias (BIAS) of WRF-Hydro
    with reference to HyDAT for discharge between May~31 (dashed line)
    and October~30.  Blue shading (abscissa) denotes a backwater ice
    impact on HyDAT water level.}
  \label{fig05}
\end{center}
\end{figure}

Before seeking a representation of {\it natural and regulated flow}
for individual catchments using neural networks, we first sought to
extend the calibration of natural flow for all catchments uniformly.
Starting with the PEST parameters obtained above, we considered a
calibration to address strong peaks in the seasonal and daily flows of
the 25~RHBN stations (Fig.~\ref{fig04}).  The NN232 neural network
(using daily and 12-day average streamflow) was trained to associate
WRF-Hydro with the 3699~RHBN observations.  Table~4 reveals
that even using NN232 streamflow, only two of 25~stations were
considered satisfactory \citep{Moriasi_etal_2015}, but as expected,
NN232 also yielded adjustments that smoothed, and in part buffered,
streamflow where peaks in the daily and 12-day averages were
different.  At the Attawapiskat River station (Fig.~\ref{fig05}), PEST
discharge (purple) was more similar to HyDAT (black) than default
WRF-Hydro parameters (orange), and the largest flows were further
reduced by NN232 (grey), which was more similar by design.

\newpage
\end{multicols}
\begin{table}
  \begin{center}
  {\hspace{1.6in}
  \begin{tabular}{ccccc}
  \hline
  Parameter & Satisfactory &  DIFF  &   NSE   &  COR \\
  Setting   &   Stations   &  (\%)  &         &      \\
  \hline
  Default   &       2      &   50   &  -0.86  & 0.69 \\
  PEST      &       2 (13) &   57   &  -0.60  & 0.70 \\
  NN232     &       2 (19) &   32   &   0.24  & 0.74 \\
  \hline
  \end{tabular}
  \hspace{1.6in}}
  \caption{Summary of the similarity between HyDAT and WRF-Hydro
    streamflow using default parameters, PEST parameters, and PEST
    parameters with neural network (NN232) post-processing (i.e.,
    training and testing employ the same reference).  Averages over
    25~stations are given for the absolute value of model-observation
    difference (DIFF, as a percent of observed streamflow),
    Nash-Sutcliffe efficiency (NSE), and Pearson correlation (COR;
    see Section~3.e for definitions).  Included are the number of
    satisfactory stations (NSE$>$0.5 and DIFF$<$15\%) and number of
    stations with improved NSE relative to default (in brackets).}
  \end{center}
\label{tab04}
\end{table}

\begin{multicols}{2}

\begin{figure}
\begin{center}
  \hbox{\includegraphics[width=0.49\textwidth]{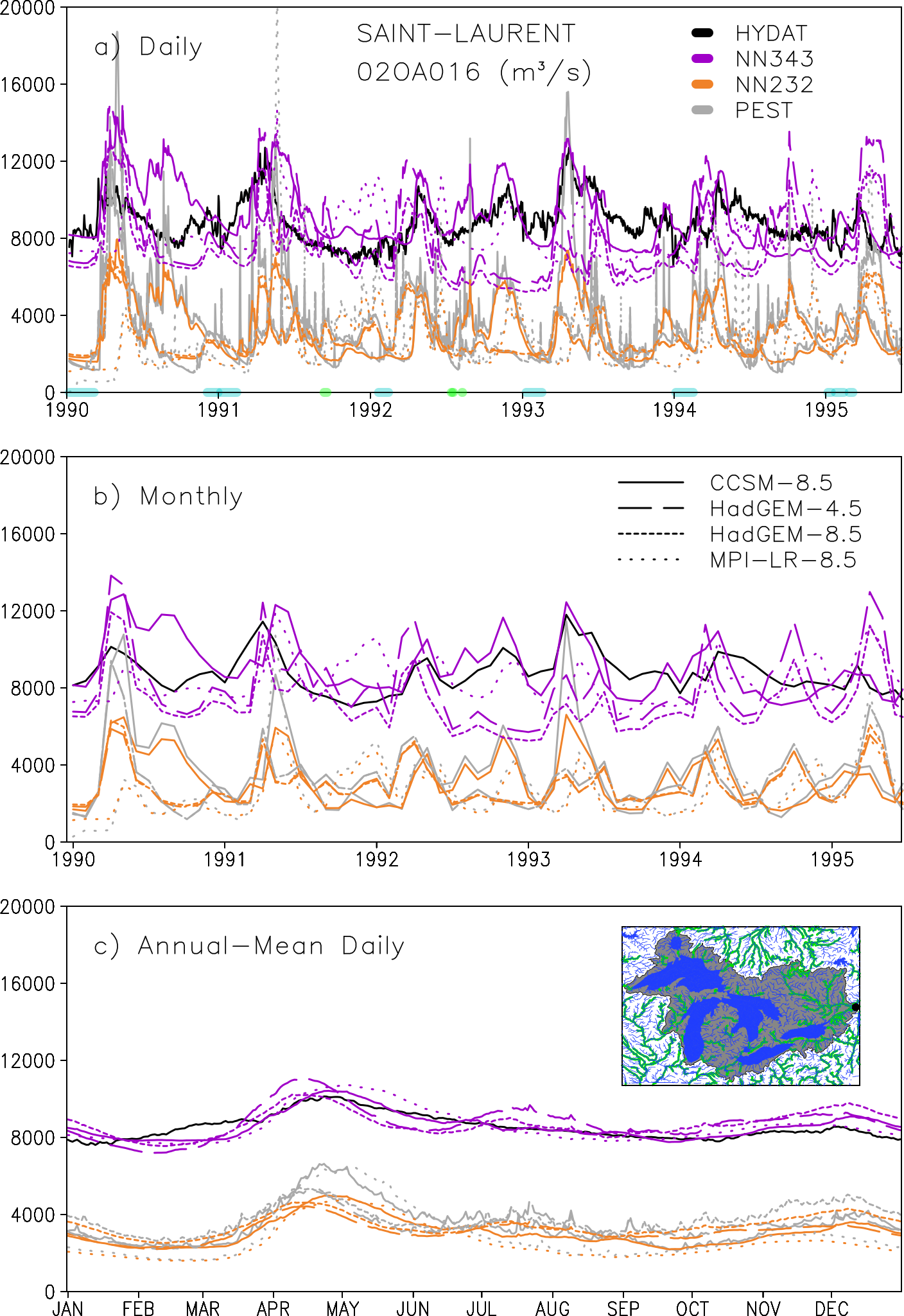}
        \hspace{0.05cm}
        \includegraphics[width=0.49\textwidth]{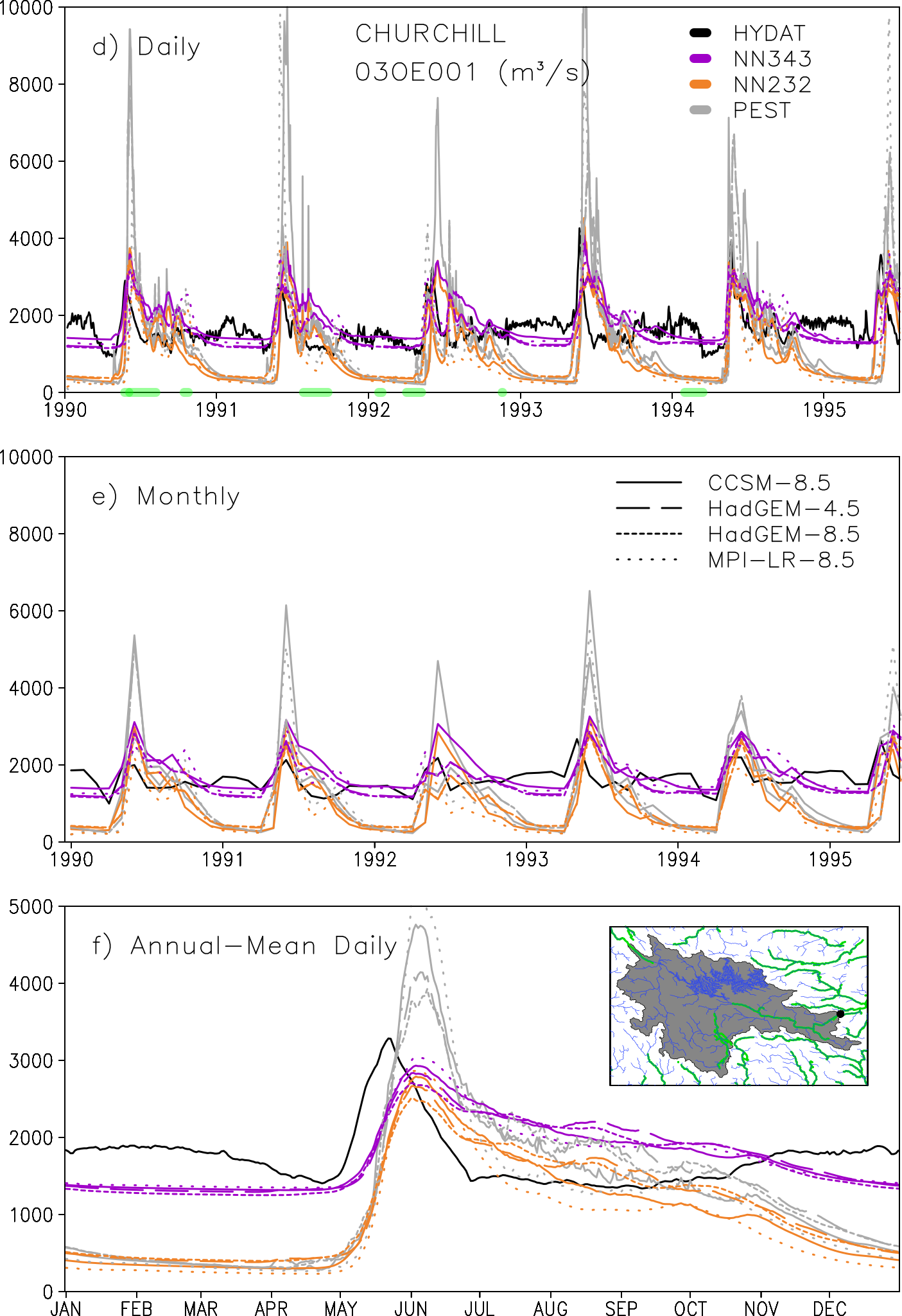}}
  \caption{Regulated discharge (m$^3$~s$^{-1}$) of the St.~Lawrence
    River at Lasalle (02OA016, left panels) and the Churchill River
    above Upper Muskrat Falls (03OE001, right panels) at a,d)~daily
    and b,e)~monthly intervals during 1990-1995, and c,f)~as annual
    averages for each day of the year between 1990 and 2022.  Shown
    are the HyDAT (black), PEST (grey), NN232 (orange), and NN343
    (purple) representations for the CCSM-4 SSP5-8.5 (solid), HadGEM2
    SSP2-4.5 (long dash) and SSP5-8.5 (short dash), and MPI-ESM1.2-LR
    SSP5-8.5 (dotted) simulations.  Insets (c,f) depict station
    location (black dot) and upstream catchment (grey shading), with
    WRF-Hydro rivers and lakes (blue) and wide rivers (green;
    \citealt{Altenau_etal_2021}).  Light blue and green shading
    (a,d~abscissae) denote an ice impact on, or estimate of, HyDAT
    water level, respectively.}
  \label{fig06}
\end{center}
\end{figure}

As multiple neural networks can be chained together
\citep{Stiles_etal_2014}, we used NN232 as a baseline for the NN343
calibration of individual catchments.  Many large rivers in eastern
Canada are regulated, and regardless of atmospheric forcing, the need
for separate seasonal calibrations is apparent.  Regulated flows of
the St.~Lawrence and Churchill Rivers are shown in Fig.~\ref{fig06}.
Spring peaks in WRF-Hydro (grey) often exceed those in HyDAT (black),
and as expected, regulated HyDAT flow is otherwise larger for the
remainder of the year.  Similarity among CMIP simulations and their
systematic differences with HyDAT are more apparent in the annual
averages (Fig.~\ref{fig06}c,f).  Neural network calibration is shown
here to demonstrate that the largest flows (NN232, orange lines) and
seasonal differences (NN343, purple lines) are reduced, although
adjustments at both stations are incomplete.  That is, HyDAT station
02OA016 is a partial measure of St.~Lawrence streamflow
\citep{Morin_Champoux_2006}, and for Churchill, NN343 is slow to
capture the flow increase in early May and a return to low flow in
July-October.

For all remaining WRF-Hydro simulations, we adjusted watershed
boundaries upstream of the 477~ocean outlets (Section~3.a) and
included 20~lakes larger than 1000~km$^2$ \citep{Messager_etal_2016}.
Although lake shapefiles were also adjusted to match the 2-km river
routing grid, inside these lakes we replaced time-invariant surface
characteristics (in geogrid.nc) with adjacent land values.  This was
done to ensure the stability of long-running simulations, but
contributed to systematic differences inherited by the NN343
calibration, with low flow in the St.~Lawrence (Fig.~\ref{fig06}c)
being an obvious example.

\newpage
\end{multicols}
\begin{multicols}{2}

\begin{figure}
\begin{center}
  \includegraphics[width=0.80\textwidth]{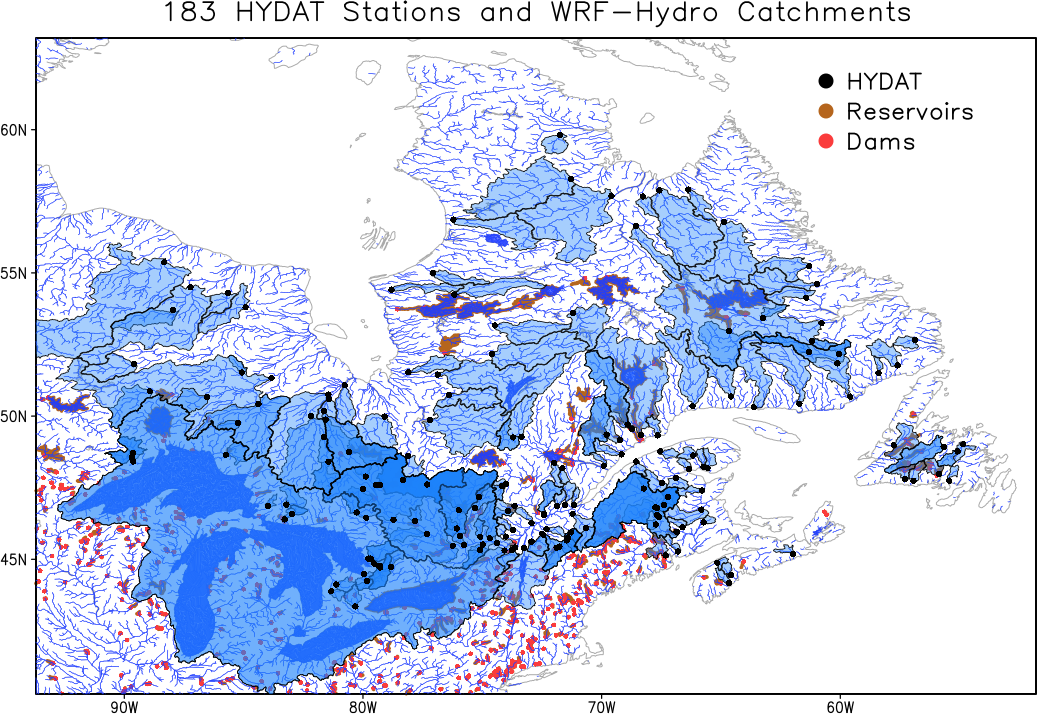}
  \caption{Catchments (overlapping light blue shading) upstream of
    183~HyDAT stations (black dots), where 1236444~daily observations
    are available between 1990 and 2022.  Also shown is the 2-km
    WRF-Hydro river and lake network (dark blue), and reservoirs
    (brown) and dams (red), as given by \citet{Lehner_etal_2011}.}
  \label{fig07}
\end{center}
\end{figure}

Training and testing of NN343 employed ERA5 forcing (1990-2022), and
if gains in similarity to HyDAT were obtained, then NN343 was
retrained for each CMIP forcing (i.e., with daily streamflow matched
by rank).  Ultimately, we applied NN343 at about 10\% (51/477) of the
ocean outlets in Fig.~\ref{fig01}, but with ERA5 forcing, we included
183~HyDAT stations with at least 1000~streamflow observations between
1990 and 2022.  Some catchments overlapped (Fig.~\ref{fig07}) and this
did not include stations whose upstream catchment area on the
WRF-Hydro and HydroSHEDS routing grids differed by more than a factor
of 1.2.  Two NN343 calibrations were trained and tested in parallel,
by splitting 604449~observations from even years and
631995~observations from odd years (i.e., trained using one set and
tested using the other).

\newpage
\end{multicols}

\begin{table}
  \begin{center}
  {\begin{tabular}{ccccccccc}
  \hline
  Satisfac.       &      DIFF     &        NSE        &         COR          & Avg. &    Satisfac.    &      DIFF     &        NSE        &          COR         \\
  Stations        &      (\%)     &                   &                      &      &    Stations     &      (\%)     &                   &                      \\
  \hline
  14/41/{\bf  84} & 61/46/{\bf 9} & -9/-3/{\bf 0.48}  & 0.58/0.62/{\bf 0.68} & Day  & 10/43/{\bf  90} & 62/46/{\bf 8} &  -8/-3/{\bf 0.43} & 0.58/0.62/{\bf 0.66} \\
  37/52/{\bf 113} & 61/46/{\bf 9} & -9/-3/{\bf 0.57}  & 0.69/0.68/{\bf 0.75} & Mon. & 40/55/{\bf 124} & 62/46/{\bf 8} &  -7/-3/{\bf 0.53} & 0.69/0.69/{\bf 0.74} \\
  33/51/{\bf 112} & 61/46/{\bf 9} & -7/-3/{\bf 0.62}  & 0.67/0.69/{\bf 0.76} & Ann. & 34/56/{\bf 125} & 62/46/{\bf 8} & -10/-4/{\bf 0.57} & 0.68/0.70/{\bf 0.74} \\
  \hline
  \end{tabular}}
  \caption{Similarity of ERA5-forced streamflow to HyDAT, as in
    Table~4, but for the 183~stations of Fig.~\ref{fig07}
    after calculating daily, monthly, and annual-mean (daily) averages.
    Neural network training and testing employ even- and odd-year
    streamflow (leftmost columns), and odd- and even-year streamflow
    (rightmost columns), respectively.  Values are shown for
    PEST/NN232/NN343 and bold denotes the calibration with greatest
    similarity between WRF-Hydro and HyDAT.}
  \end{center}
\label{tab05}
\end{table}
\begin{multicols}{2}

\begin{figure}
\begin{center}
  \includegraphics[width=0.70\textwidth]{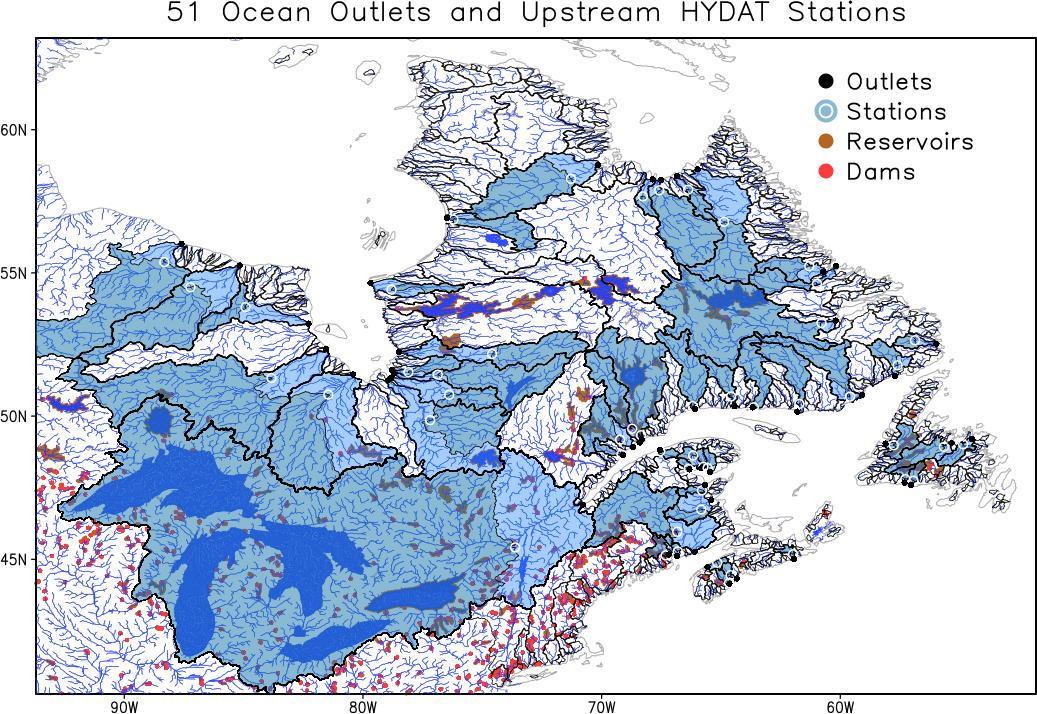}
  \caption{As in Fig.~\ref{fig07}, but for 51~ocean outlets (black
    dots) whose nearest inland HyDAT stations (white circles) have
    upstream catchment areas (grey overlap) that are at least 40\% of
    the ocean outlet catchment areas (light blue).  A total of
    366743~daily observations are available between 1990 and 2022.}
  \label{fig08}
\end{center}
\end{figure}

Table~5 confirms the NN343 gains in similarity, with
odd-year (left columns) and even-year (right columns) test results
being nearly equivalent.  A convenient summary is given by the number
of satisfactory stations \citep{Moriasi_etal_2015}.  With PEST and
NN232 calibration, we found between 41 and 56 stations (22\% to 31\%)
to be satisfactory, and by including the NN343 calibration, station
number more than doubled (46\% to 68\%; bold values).  This prompted a
second attempt to find a uniform calibration for all natural flows
across eastern Canada (gauged or not), but unfortunately, we found no
average gain when training a single NN343 calibration simultaneously
at a group of 85~HyDAT stations (not shown).  Thus, NN343 training
focused on individual HyDAT stations, and specifically, any whose
upstream catchment covered at least 40\% of the corresponding ocean
outlet catchment (Fig.~\ref{fig08}).  This yielded 51~HyDAT stations
that we could use to calibrate ocean outlets downstream
\citep{Dai_Trenberth_2002}.

\newpage
\end{multicols}

\begin{table}
  \begin{center}
  {\begin{tabular}{ccccccccc}
  \hline
  Satisfac.      &      DIFF     &       NSE       &         COR          & Avg. &    Satisfac.    &      DIFF     &       NSE        &          COR         \\
  Stations       &      (\%)     &                 &                      &      &    Stations     &      (\%)     &                  &                      \\
  \hline
   4/15/{\bf 26} & 30/20/{\bf 7} & -1/0/{\bf 0.4}  & 0.62/0.68/0.68       & Day  &  4/17/{\bf 27} & 29/20/{\bf 8} & -1/ 0/{\bf 0.4} & 0.62/0.69/0.69       \\
  10/18/{\bf 33} & 29/20/{\bf 6} & -1/0/{\bf 0.5}  & 0.75/0.75/{\bf 0.76} & Mon. & 14/19/{\bf 35} & 29/20/{\bf 8} & -1/ 0/{\bf 0.5} & 0.76/0.76/{\bf 0.77} \\
  10/18/{\bf 36} & 28/19/{\bf 7} & -1/0/{\bf 0.5}  & 0.72/0.75/0.75       & Ann. & 12/20/{\bf 35} & 29/19/{\bf 8} & -2/-1/{\bf 0.5} & 0.75/0.78/0.78       \\
  \hline
  \end{tabular}}
  \caption{As in Table~5, but for the 51~HyDAT stations of
    Fig.~\ref{fig08}.  Representations shown are the
    PEST/NN232/NN343-seasonal values, where the latter is a seasonal
    CMIP calibration (see text).}
  \end{center}
\label{tab06}
\end{table}
\begin{multicols}{2}

We performed one final test of NN343 to examine the impact of
streamflow matching by rank (Section~3.d).  This employed ERA5 forcing
and the 51~HyDAT stations, and during training, the daily streamflow
of WRF-Hydro was matched by rank to that of HyDAT, while the order of
monthly and annual-mean values were unchanged.  In spite of this
partial resorting, gains in HyDAT similarity (Table~6) were
comparable to gains for all 183~stations (Table~5), with
over half of the 51~stations being satisfactory.

\begin{figure}[hp]
\begin{center}
  \includegraphics[width=0.87\textwidth]{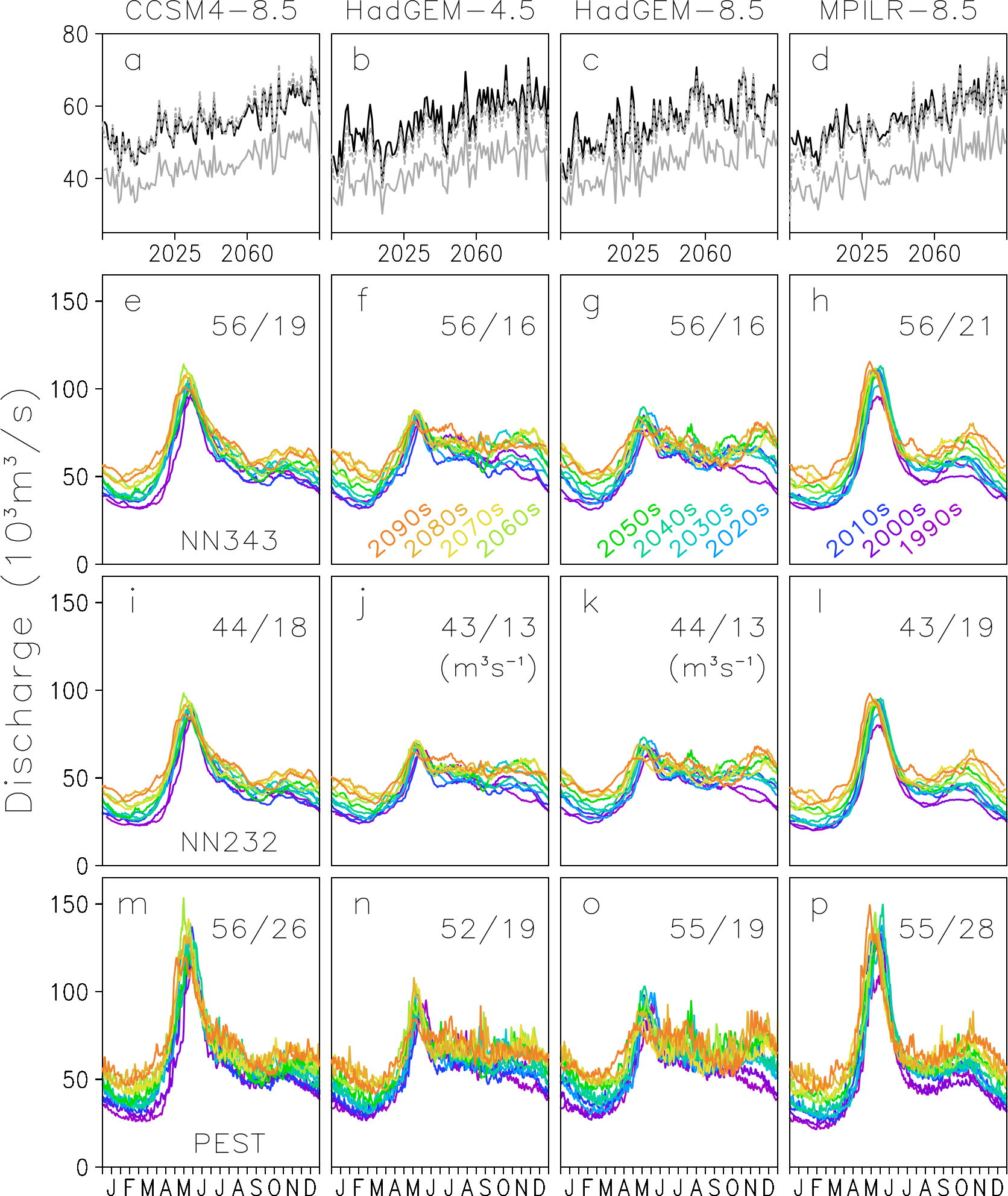}
  \caption{Simulated 1990-2100 discharge ($10^3$~m$^3$~s$^{-1}$) for
    the 477~eastern Canadian rivers between the Severn and St.~Croix
    (Fig.~\ref{fig01}) using downscaled CMIP forcing from CCSM-4
    SSP5-8.5 (left column), HadGEM2 SSP2-4.5 and SSP5-8.5 (center-left
    and right), and MPI-ESM1.2-LR SSP5-8.5 (right column).  Shown are
    a-d)~annual averages and e-p)~daily averages for decades (colours)
    for NN343 (e-h; a-d grey dashed lines), NN232 (i-l; a-d grey
    lines), and PEST (m-p; a-d black lines) calibrations.  Panels
    (e-p) include mean/standard deviation of daily discharge at upper
    right.}
  \label{fig09}
\end{center}
\end{figure}

All steps in Table~1 were taken to produce the four
calibrated projections of eastern Canadian discharge.  For each of the
CCSM-4, HadGEM2, and MPI-ESM1.2-LR models, a three-year WRF-Hydro
spinup was followed by simulations from 1990 to 2100 using PEST
parameters, with the NN232 calibration at all 477~ocean outlets, and a
NN343 calibration at 51~outlets.  The discharge from all rivers
between the Severn and St.~Croix was then summed, and Figure~\ref{fig09}
shows the annual and decadal averages, with peak discharge in May of
about 10$^5$~m$^3$~s$^{-1}$.  As expected, each simulation captured
increasing annual discharge (Fig.~\ref{fig09}a-d), and as others have
found (e.g., \citealt{Bush_Lemmen_Bonsal_etal_2019,
  Stadnyk_etal_2021}), the decadal averages revealed increasing low
flow during the cold season and an earlier peak discharge in spring
(Fig.~\ref{fig09}e-p).  The calibration of each CMIP simulation to a
common set of observational references was expected to yield greater
precision, and this was also confirmed.  Without calibration, the
daily discharge mean and standard deviation are more disparate across
the four models (Fig.~\ref{fig09}m-p), whereas with calibration, the
CMIP simulations share a mean of $56 \times 10^3$~m$^3$~s$^{-1}$ and a
more similar variance (Fig.~\ref{fig09}e-h).  There are also striking
differences among the four discharge simulations, even after
calibration.  However, attribution of the reduced seasonality of the
HadGEM2 simulations in a hydroclimatic assessment that includes CCSM-4
and MPI-ESM1.2-LR is beyond our present scope.

\newpage
\end{multicols}
\begin{multicols}{2}

\section{Discussion}

This climate configuration of WRF-Hydro employs lower spatiotemporal
resolution, but after calibration to historical records
\citep{Pellerin_NzokouTanekou_2020, deRham_etal_2020, Hydat_2023}, our
streamflow predictions compare favourably to similar high latitude
modelling efforts.  For instance, the spatially uniform calibration
given by PEST and NN232 yields monthly gains in similarity on par with
those of \citet{Stadnyk_etal_2021}, who found 30\% of stations to be
satisfactory.  Moreover, at catchments where an individual NN343
calibration could also be performed, satisfactory stations more than
doubled to 62\%-68\% (Tables~5 and~6).

We treat NN232 and NN343 as extensions of physical models that address
processes like streamflow buffering and regulation, but unlike the
nine WRF-Hydro parameters of Table~2, we do not associate
any of the 23 or 39 parameters with specific processes because our
densely connected structures are ad hoc.  When physical processes are
not apparent, neural networks are adaptive and convenient, but their
use seems justifiable because they do well at {\it parameterizing}
nonlinear associations (Section~3.d).  For instance, the 1990-2100
WRF-Hydro trends in Fig.~\ref{fig09}a-d appear to be shifted after
each calibration step by a constant value.  On one hand, this means
that calibrated trends are well behaved \citep{Maraun_2016} and an
analysis of the uncalibrated trends would be equivalent
(cf.~\citealt{Stadnyk_etal_2021}).  On the other hand,
\citet{Stadnyk_etal_2021} point out that regulation is expected to
have an impact on trends.  Thus, we would caution that a)~relatively
simple neural networks like NN232 and NN343 seem to provide well
behaved process parameterizations for 1990-2022, and b)~we are making
the conventional assumption that such parameterizations can be applied
well beyond 2022 \citep{Maraun_2016}.

We emphasize hydrologic model output (ocean forcing) in this study,
but equal emphasis on the input (atmospheric forcing) is needed in an
Earth system modelling context.  In lieu of our simple adjustment of
atmospheric forcing (Section~3.b), either a neural network calibration
(Section~3.d) or bias correction \citep{Cannon_2018} can be
considered.  Moreover, in lieu of neural network calibration of ocean
forcing (streamflow), bias correction can also be considered.  Seeking
to preserve intervariable dependence is advantageous
\citep{Cannon_2018}, but bias correction also reorders slightly the
sequence of synoptic events, which can shorten their duration
\citep{Cannon_2016}.  No attempt was made to preserve intervariable
dependence here.  Although NN343 also reorders daily streamflow (i.e.,
by matching rank), this is only done during training, so unlike bias
correction, its input and output time sequence is the same.  Following
\citet{Cannon_2018}, both methods preserve trends of increasing annual
discharge (Fig.~\ref{fig09}a-d) and seem well behaved
\citep{Maraun_2016}.

A balanced emphasis on the internal calibration of WRF-Hydro also has
merit, yet we opted not to allow PEST to adjust more than nine
WRF-Hydro parameters, including Manning's roughness coefficients for
rivers (cf.~\citealt{Wang_etal_2019}), or to allow the WRF-Hydro
parameters to vary spatially \citep{Mendoza_etal_2015,
  RafieeiNasab_etal_2020, Gochis_etal_2021}.  However, each
calibration depends on those before it, and we expect that NN232 and
NN343 compensate in part for the tuning that could be done internally
to WRF-Hydro, say, to represent regulation and buffering by small
lakes \citep{Dery_etal_2018, Stadnyk_etal_2020}.  In this study, we
also address a less obvious compensation between data and process
models, where it is common to develop parameterizations for WRF-Hydro
(or like NN343), and less common to develop the data models that we
use to select their parameterization values, like by (\ref{optp}) and
(\ref{optn}).  In spite of the seemingly different calibration steps
that we employ, there may be some benefit to expressing data models in
a way that allows their fundamental similarites (Table~1)
and differences \citep{Beven_2021} to be recognized within a common
and familiar framework (Appendix).

\section{Conclusions}

A calibration of WRF-Hydro with respect to eastern Canadian streamflow
observations from 1990-2022 is performed in support of regional
hydroclimate simulations through 2100.  The WRF-Hydro model
\citep{Gochis_etal_2021} is applicable over a wide range of
spatiotemporal scales and benefits from an efficient automatic
calibration \citep{Doherty_2015, Wang_etal_2019} of its model
parameters \citep{RafieeiNasab_etal_2020}.  Starting with a relatively
low resolution (50-km/2-km) grid and (6-h ERA5) atmospheric forcing,
we modify watershed boundaries of the \citet{Eilander_etal_2021}
digital elevation model to more easily relate predicted and observed
streamflow at gauging stations.  Following \citet{Doherty_2015}, we
then use PEST to identify nine spatially invariant WRF-Hydro
parameters.  Because some parameter estimates are close to a limit in
their range, however, we cannot identify a uniform model
representation of natural flow through the eastern Canadian river
network.

By way of committing to a calibration by individual (regulated)
watershed and CMIP forcing \citep{Taylor_etal_2012, Eyring_etal_2016},
we calibrate WRF-Hydro streamflow using a pair of neural networks
(NN232 and NN343) that build on the PEST calibration.  First, gains in
similarity to HyDAT streamflow observations
\citep{Pellerin_NzokouTanekou_2020, deRham_etal_2020, Hydat_2023} are
confirmed for NN232, which is trained to reduce strong peaks in
seasonal and daily discharge simulations.  With over $10^6$ streamflow
observations during 1990-2022, gains in similarity are then confirmed
using PEST, NN232, and NN343 calibration, where NN343 is trained at
183~HyDAT stations.  An experiment to train NN343 jointly at
85~stations does not yield an average gain in similarity.

Training of NN343 is performed separately for four CMIP-forcing
simulations and at 51~HyDAT stations that permit a downstream
adjustment of discharge, following \citet{Dai_Trenberth_2002}.  This
training is novel in that CMIP simulations are calibrated directly to
HyDAT observations, as NN343 employs three input and output nodes.
Similar to \citet{Freilich_Challenor_1994}, training employs a sorting
of WRF-Hydro and HyDAT streamflow that are matched by rank, but only
for the daily input/output node.  A seasonal calibration is thus
obtained by not sorting at the other two nodes (i.e., the monthly and
annual averages).  This yields gains in similarity at the HyDAT
stations that are equivalent to those of 183~stations, with over half
the 51~stations being satisfactory \citep{Moriasi_etal_2015,
  Stadnyk_etal_2021}.

The resulting hydrologic data-calibration steps are applied to obtain
eastern Canadian freshwater discharge through the 477~WRF-Hydro ocean
outlets for each of the four CMIP models.  Whereas WRF-Hydro with PEST
parameters yields strong seasonal and daily peaks relative to HyDAT,
and these peaks are smoothed by NN232, the separate NN343 calibration
at 51~ocean outlets yields relatively precise annual variations and
interannual trends for all CMIP forcings.  This provides further
support for an increasing low flows during future cold seasons and an
earlier peak discharge in future spring seasons
\citep{Bush_Lemmen_Bonsal_etal_2019, Stadnyk_etal_2021}.

\section*{Acknowledgements}

We thank Adam Drozdowski, Zhenxia Long, Bash Toulany, and two
anonymous reviewers who provided thoughtful comments and a
significantly updated presentation structure.  Funding was provided by
the Competitive Science Research Fund (CSRF) of Fisheries and Oceans
Canada (DFO).  All models and data have been freely provided, and we
thank the WRF-Hydro (NCAR), MERIT (University of Tokyo, Vrije
Universiteit Amsterdam), HydroSHEDS (WWF, McGill University), CMIP
(World Climate Research Programme), HyDAT and RHBN (ECCC) model and
data development teams.  DFO and the Digital Research Alliance of
Canada provided computational resources and support.  Data processing
and visualization employed Julia (Julia Computing Inc., MIT),
WhiteBoxTools (University of Guelph), and GrADs (COLA, George Mason
University), respectively.  Gitlab hosts code shared by the DFO
modelling community (MODCOM), including our configuration steps
(https://gitlab.com/dfo\_modcom/diag.hydrology).

\section*{Appendix -- A data modelling primer}

There are active discussions in the Earth sciences about what we might
describe as data modelling within the familiar context of process
modelling.  Here, we use data models to provide a mathematical
framework for calibration and confirmation, including the definition
of an ``equivalence assumption'' and supporting terms.  To acknowledge
that data modelling and process modelling may be {\it complementary},
we specify data while not being specific about the hydrologic
processes that they refer to.  The common notions of association that
we sometimes express in words can often be expressed as a data model.
In turn, we claim that data models provide a basis for calibration and
confirmation, even if it might not be easy to recognize such a
foundation within a familiar knowledge system \citep{Beven_2021,
  Nearing_etal_2016, Kalman_1994}.  For example, the expression ``this
watershed boundary is our reference'' can be written as a common
model, as can all associations that are made in this study.

At each step toward what is observed and expected, we make a
particular representation (i.e., predictions or observations) more
similar to a selected reference (i.e., an ideal representation) using
a data model of the form
\begin{equation}
  \tag{3}
  C = t + \epsilon_c \mbox{\hspace{0.3in} and \hspace{0.3in}} U = t + \epsilon_u.
%  \nonumber
  \label{errinvar}
\end{equation}
Here, $C$ and $U$ are the calibrated and uncalibrated representations
(e.g., observed and predicted streamflow) with errors $\epsilon_c$ and
$\epsilon_u$, respectively.  We refer to (\ref{errinvar}) as a target
model if the calibration reference ($C$) includes error, and
otherwise, as a truth model if $\epsilon_c = 0$, and a truth-truth
model if $\epsilon_u = 0$ as well.

\vspace{0.5cm}
\noindent {\it Glossary of model terms}
\vspace{0.5cm}

{\bf Bias correction}: This is an association of two representations,
expressed in terms of the mean, variance, or quantiles of $C$ and $U$
\citep{Ehret_etal_2012, Maraun_2016}.  Systematic differences in the
representations of one Earth system model component and another (or
that of a regional model or local observations) generally involve
multivariate associations between $C$ and $U$ \citep{Cannon_2018}, and
by the truth-truth model, their covariance and quantiles are taken as
error free \citep{Cannon_2016}.

{\bf Data model}: We acknowledge the need to explore different types
of associations within a mathematical knowledge system by, for
example, including the notion of epistemic observational uncertainty
\citep{Beven_2021}.  Clearly, a general framework is desirable
\citep{Kalman_1994, Nearing_etal_2016} and it seems worthwhile to
consider common expressions like (\ref{errinvar}) as contributing to
such a framework (i.e., where data includes predictions and
observations).  {\it Even when no calculations are performed}
(Section~3.a,b), a data model provides a mathematical foundation for
calibration, as the relationship between predictions and observations
assumes such a model, and this can guide a formulation of measurement
performance, which in turn, can guide our predictive model adjustments.

{\bf Equivalence assumption}: Data models may be common, on one hand,
but it is important to note that while $C$ and $U$ are related via
$t$, they are expressed separately in (\ref{errinvar}) because each
representation {\it of nature} is expected to capture processes
somewhat differently \citep{Lorenz_1985, Smith_2006, Parker_2017b,
  Beven_2021}.  On the other hand, the use of (\ref{errinvar}) in
parameter estimation \citep{Doherty_2015}, data assimilation
\citep{Mahfouf_1991}, machine learning \citep{Kratzert_etal_2019a},
and model confirmation \citep{Moriasi_etal_2015} typically follows
from equating $C$ and $U$ in order to minimize the model-observation
difference.  We employ this assumption thoughout our study.

{\bf Knowledge systems}: We acknowledge the association of other
knowledge systems with the scientific, mathematical, and linguistic
systems that we relied on to explore hydrologic aspects in this study.
We also acknowledge that these aspects of nature are expressed in
knowledge systems other than the ones we are familiar with.

{\bf Reference}: Any dataset (predictions or observations) might be
considered a useful depiction of some aspect of nature.  It is
convenient in this study to take one representation (the reference) as
a temporary proxy of nature, but caution is warranted, as an
association of two datasets might only be linear where there is an
overlap in their representations (e.g., processes resolved by both).
In taking one as a reference, only nonlinear association might be
possible where representations do not overlap (e.g., via interactions
with unresolved processes; Section~3.d).

{\bf Representation}: Any dataset (predictions or observations) might
be considered an improving depiction of some aspect of nature.  A
representation is such a depiction (e.g., water level at a gauging
station or multivariate data on the WRF-Hydro grid).  Even if a
familiar representation is the result of many years of development
with associations to many datasets, strictly from a metrological point
of view, this remains a specific representation with specific
associations \citep{Bulgin_etal_2022}.

\end{multicols}

%\bibliographystyle{apacite}
%\begin{thebibliography}{}
%\end{thebibliography}
\bibliographystyle{amets}
%\bibliography{refer}

\begin{thebibliography}{59}
\expandafter\ifx\csname natexlab\endcsname\relax\def\natexlab#1{#1}\fi

\bibitem[Altenau et~al.(2021)Altenau, Pavelsky, Durand, Yang, Frasson, and
  Bendezu]{Altenau_etal_2021}
Altenau, E.~H., T.~M. Pavelsky, M.~T. Durand, X.~Yang, R.~P. d.~M. Frasson, and
  L.~Bendezu, 2021: {The Surface Water and Ocean Topography (SWOT) Mission
  River Database (SWORD): A global river network for satellite data products}.
  {\em Water Resour. Res.\/}, {\bf 57}, 1--15, doi:10.1029/2021WR030054.

\bibitem[Berg et~al.(2018)Berg, Donnelly, and Gustafsson]{Berg_etal_2018}
Berg, P., C.~Donnelly, and D.~Gustafsson, 2018: {Near-real-time adjusted
  reanalysis forcing data for hydrology}. {\em Hydrol. Earth System Sci.\/},
  {\bf 22}, 989--1000, doi:10.5194/hess--22--989--2018.

\bibitem[Beven(2019)]{Beven_2019}
Beven, K., 2019: {Towards a methodology for testing models as hypotheses in the
  inexact sciences}. {\em Proc. Roy. Soc. A\/}, {\bf 475}, 1--19,
  doi:10.1098/rspa.2018.0862.

\bibitem[Beven(2021)]{Beven_2021}
Beven, K., 2021: {An epistemically uncertain walk through the rather fuzzy
  subject of observation and model uncertainties}. {\em Hydrol. Processes\/},
  {\bf 35}, 1--9, doi:10.1002/hyp.14012.

\bibitem[Bonsal et~al.(2019)Bonsal, Peters, Seglenieks, Rivera, and
  Berg]{Bush_Lemmen_Bonsal_etal_2019}
Bonsal, B.~R., D.~L. Peters, F.~Seglenieks, A.~Rivera, and A.~Berg, 2019:
  \textnormal{{Changes in freshwater availability across Canada, Chapter~6 of
  {\it Canada's Changing Climate Report}, E. Bush and D. S. Lemmen, Eds.,
  Government of Canada, Ottawa, Ontario, p.~261--342, (accessed April 2024 at
  https://changingclimate.ca/CCCR2019)}}.

\bibitem[Bulgin et~al.(2022)Bulgin, Thomas, Waller, and
  Woolliams]{Bulgin_etal_2022}
Bulgin, C.~E., C.~M. Thomas, J.~A. Waller, and E.~R. Woolliams, 2022:
  {Representation uncertainty in the Earth sciences}. {\em Earth Space Sci.\/},
  {\bf 9}, 1--7, doi:10.1029/2021EA002129.

\bibitem[Bush and Lemmen(2019)]{Bush_Lemmen_etal_2019}
Bush, E., and D.~S. Lemmen, 2019: \textnormal{{Canada's Changing Climate
  Report, Government of Canada, Ottawa, Ontario, 444 pp., (accessed April 2024
  at https://changingclimate.ca/CCCR2019)}}.

\bibitem[Cannon(2016)]{Cannon_2016}
Cannon, A.~J., 2016: {Multivariate bias correction of climate model output:
  Matching marginal distributions and intervariable dependence structure}. {\em
  J. Climate\/}, {\bf 29}, 7045--7064, doi:10.1175/JCLI--D--15--0679.1.

\bibitem[Cannon(2018)]{Cannon_2018}
Cannon, A.~J., 2018: {Multivariate quantile mapping bias correction: An
  N-dimensional probability density function transform for climate model
  simulations of multiple variables}. {\em Climate Dyn.\/}, {\bf 50}, 31--49,
  doi:10.1007/s00382--017--3580--6.

\bibitem[Clevert et~al.(2015)Clevert, Unterthiner, and
  Hochreiter]{Clevert_etal_2015}
Clevert, D.-A., T.~Unterthiner, and S.~Hochreiter, 2015: \textnormal{{Fast and
  accurate deep network learning by exponential linear units (ELUs),
  arXiv:1511.07289 [cs.LG]}}.

\bibitem[Collins et~al.(2011)Collins, Bellouin, Doutriaux-Boucher, Gedney,
  Halloran, Hinton, Hughes, Jones, Joshi, Liddicoat, Martin, O’Connor, Rae,
  Senior, Sitch, Totterdell, Wiltshire, and Woodward]{Collins_etal_2011}
Collins, W.~J., N.~Bellouin, M.~Doutriaux-Boucher, N.~Gedney, P.~Halloran,
  T.~Hinton, J.~Hughes, C.~D. Jones, M.~Joshi, S.~Liddicoat, G.~Martin,
  F.~O’Connor, J.~Rae, C.~Senior, S.~Sitch, I.~Totterdell, A.~Wiltshire, and
  S.~Woodward, 2011: Development and evaluation of an {Earth-System model --
  HadGEM2}. {\em Geosci. Model Dev.\/}, {\bf 4}, 1051--1075,
  doi:10.5194/gmd--4--1051--2011.

\bibitem[Dai and Trenberth(2002)]{Dai_Trenberth_2002}
Dai, A., and K.~E. Trenberth, 2002: {{Estimates of freshwater discharge from
  continents: Latitudinal and seasonal variations}}. {\em J. Hydrometeor.\/},
  {\bf 3}, 660--687, doi:10.1175/1525\--7541(2002)003$<$0660:EOFDFC$>$2.0.CO;2.

\bibitem[Dai et~al.(2009)Dai, Qian, Trenberth, and Milliman]{Dai_etal_2009}
Dai, A., T.~Qian, K.~E. Trenberth, and J.~D. Milliman, 2009: {{Changes in
  continental freshwater discharge from 1948 to 2004}}. {\em J. Climate\/},
  {\bf 22}, 2773--2792, doi:10.1175/2008JCLI2592.1.

\bibitem[{de Rham} et~al.(2020){de Rham}, Dibike, Beltaos, Peters, Bonsal, and
  Prowse]{deRham_etal_2020}
{de Rham}, L., Y.~Dibike, S.~Beltaos, D.~Peters, B.~Bonsal, and T.~Prowse,
  2020: {A Canadian river ice database from the National Hydrometric Program
  archives}. {\em Earth Syst. Sci. Data\/}, {\bf 12}, 1835--1860,
  doi:10.5194/essd--12--1835--2020.

\bibitem[Dee(2005)]{Dee_2005}
Dee, D.~P., 2005: Bias and data assimilation. {\em Quart. J. Roy. Meteor.
  Soc.\/}, {\bf 131}, 3323--3343, doi:10.1256/qj.05.137.

\bibitem[Derksen et~al.(2019)Derksen, Burgess, Duguay, Howell, Mudryk, Smith,
  Thackeray, and Kirchmeier-Young]{Bush_Lemmen_Derksen_etal_2019}
Derksen, C., D.~Burgess, C.~Duguay, S.~Howell, L.~Mudryk, S.~Smith,
  C.~Thackeray, and M.~Kirchmeier-Young, 2019: \textnormal{{Changes in snow,
  ice, and permafrost across Canada, Chapter~5 of {\it Canada's Changing
  Climate Report}, E. Bush and D. S. Lemmen, Eds., Government of Canada,
  Ottawa, Ontario, p.~194--260}}.

\bibitem[D{\'e}ry et~al.(2018)D{\'e}ry, Stadnyk, MacDonald, Koenig, and
  Guay]{Dery_etal_2018}
D{\'e}ry, S.~J., T.~A. Stadnyk, M.~K. MacDonald, K.~A. Koenig, and C.~Guay,
  2018: {Flow alteration impacts on Hudson Bay river discharge}. {\em Hydrol.
  Processes\/}, {\bf 32}, 3576--3587, doi:10.1002/hyp.13285.

\bibitem[Doherty(2015)]{Doherty_2015}
Doherty, J., 2015: {\it Calibration and Uncertainty Analysis for Complex
  Environmental Models,} \textnormal{{with PEST: Model Independent Parameter
  Estimation, User Manual, 7th ed., available at https://pesthomepage.org
  (accessed May 2023), Watermark Numerical Computing, Brisbane, Australia,
  227~pp.}}

\bibitem[Ehret et~al.(2012)Ehret, Zehe, Wulfmeyer, Warrach-Sagi, and
  Liebert]{Ehret_etal_2012}
Ehret, U., E.~Zehe, V.~Wulfmeyer, K.~Warrach-Sagi, and J.~Liebert, 2012: Should
  we apply bias correction to global and regional climate model data? {\em
  Hydrol. Earth System Sci.\/}, {\bf 16}, 3391--3404,
  doi:10.5194/hess--16--3391--2012.

\bibitem[Eilander et~al.(2021)Eilander, {van Verseveld}, Yamazaki, Weerts,
  Winsemius, and Ward]{Eilander_etal_2021}
Eilander, D., W.~{van Verseveld}, D.~Yamazaki, A.~Weerts, H.~C. Winsemius, and
  P.~J. Ward, 2021: A hydrography upscaling method for scale-invariant
  parametrization of distributed hydrological models. {\em Hydrol. Earth System
  Sci.\/}, {\bf 25}, 5287--5313, doi:10.5194/hess--25--5287--2021.

\bibitem[Eyring et~al.(2016)Eyring, Bony, Meehl, Senior, Stevens, Stouffer, and
  Taylor]{Eyring_etal_2016}
Eyring, V., S.~Bony, G.~A. Meehl, C.~A. Senior, B.~Stevens, R.~J. Stouffer, and
  K.~E. Taylor, 2016: {Overview of the Coupled Model Intercomparison Project
  Phase 6 (CMIP6) experimental design and organization}. {\em Geosci. Model
  Dev.\/}, {\bf 9}, 1937--1958, doi:10.5194/gmd--9--1937--2016.

\bibitem[Flato et~al.(2019)Flato, Gillett, Arora, Cannon, and
  Anstey]{Bush_Lemmen_Flato_etal_2019}
Flato, G., N.~Gillett, V.~Arora, A.~Cannon, and J.~Anstey, 2019:
  \textnormal{{Modelling future climate change, Chapter~3 of {\it Canada's
  Changing Climate Report}, E. Bush and D. S. Lemmen, Eds., Government of
  Canada, Ottawa, Ontario, p.~74--111, (accessed April 2024 at
  https://changingclimate.ca/CCCR2019)}}.

\bibitem[Freilich and Challenor(1994)]{Freilich_Challenor_1994}
Freilich, M.~H., and P.~G. Challenor, 1994: A new approach for determining
  fully empirical altimeter wind speed model functions. {\em J. Geophys.
  Res.\/}, {\bf 99}, 25051–25062, doi:10.1029/94JC01996.

\bibitem[Gochis et~al.(2021)Gochis, Barlage, Cabell, Casali, Dugger,
  FitzGerald, McAllister, McCreight, {Rafieei Nasab}, Read, Sampson, Yates, and
  Zhang]{Gochis_etal_2021}
Gochis, D.~J., M.~Barlage, R.~Cabell, M.~Casali, A.~Dugger, K.~FitzGerald,
  M.~McAllister, J.~McCreight, A.~{Rafieei Nasab}, L.~Read, K.~Sampson,
  D.~Yates, and Y.~Zhang, 2021: \textnormal{{The NCAR WRF-Hydro modeling system
  technical description, (Version 5.2.0), NCAR Technical Note, 108 pp.,
  available at
  https://ral.ucar.edu/sites/default/files/public/projects/wrf-hydro/technical-description-user-guide/wrf-hydrov5.2technicaldescription.pdf
  (accessed October 9, 2023).}}

\bibitem[Greenan et~al.(2019)Greenan, James, Loder, Pepin, Azetsu-Scott,
  Ianson, Hamme, Gilbert, Tremblay, Wang, and
  Perrie]{Bush_Lemmen_Greenan_etal_2019}
Greenan, B. J.~W., T.~S. James, J.~W. Loder, P.~Pepin, K.~Azetsu-Scott,
  D.~Ianson, R.~C. Hamme, D.~Gilbert, J.-{\'E}. Tremblay, X.~L. Wang, and
  W.~Perrie, 2019: \textnormal{{Changes in oceans surrounding Canada, Chapter~7
  of {\it Canada's Changing Climate Report}, E. Bush and D. S. Lemmen, Eds.,
  Government of Canada, Ottawa, Ontario, p.~343--423, (accessed April 2024 at
  https://changingclimate.ca/CCCR2019)}}.

\bibitem[Hersbach et~al.(2020)Hersbach, Bell, Berrisford, Hirahara,
  Hor{\'a}nyi, Mu{\~n}oz-Sabater, Nicolas, Peubey, Radu, Schepers, Simmons,
  Soci, Abdalla, Abellan, Balsamo, Bechtold, Biavati, Bidlot, Bonavita, {De
  Chiara}, Dahlgren, Dee, Diamantakis, Dragani, Flemming, Forbes, Fuentes,
  Geer, Haimberger, Healy, Hogan, H{\'o}lm, Janiskov{\'a}, Keeley, Laloyaux,
  Lopez, Lupu, Radnoti, {de Rosnay}, Rozum, Vamborg, Villaume, and
  Th{\'e}paut]{Hersbach_etal_2020}
Hersbach, H., B.~Bell, P.~Berrisford, S.~Hirahara, A.~Hor{\'a}nyi,
  J.~Mu{\~n}oz-Sabater, J.~Nicolas, C.~Peubey, R.~Radu, D.~Schepers,
  A.~Simmons, C.~Soci, S.~Abdalla, X.~Abellan, G.~Balsamo, P.~Bechtold,
  G.~Biavati, J.~Bidlot, M.~Bonavita, G.~{De Chiara}, P.~Dahlgren, D.~Dee,
  M.~Diamantakis, R.~Dragani, J.~Flemming, R.~Forbes, M.~Fuentes, A.~Geer,
  L.~Haimberger, S.~Healy, R.~J. Hogan, E.~H{\'o}lm, M.~Janiskov{\'a},
  S.~Keeley, P.~Laloyaux, P.~Lopez, C.~Lupu, G.~Radnoti, P.~{de Rosnay},
  I.~Rozum, F.~Vamborg, S.~Villaume, and J.-N. Th{\'e}paut, 2020: {The ERA5
  global reanalysis}. {\em Quart. J. Roy. Meteor. Soc.\/}, {\bf 146},
  1999--2049, doi:10.1002/qj.3803.

\bibitem[Innes et~al.(2018)Innes, Saba, Fischer, Gandhi, Rudilosso, Joy,
  Karmali, Pal, and Shah]{Innes_etal_2018}
Innes, M., E.~Saba, K.~Fischer, D.~Gandhi, M.~C. Rudilosso, N.~M. Joy,
  T.~Karmali, A.~Pal, and V.~B. Shah, 2018: \textnormal{{Fashionable modelling
  with Flux, arXiv:1811.01457 [cs.PL]}}.

\bibitem[{IPCC}(2013)]{IPCC_2013}
{IPCC}, 2013: \textnormal{{Summary for Policymakers. In: Climate Change 2013:
  The Physical Science Basis. Contribution of Working Group~I to the Fifth
  Assessment Report of the Intergovernmenta Panel on Climate Change, T.F.
  Stocker, D. Qin, G.-K. Plattner, M. Tignor, S.K. Allen, J. Boschung, A.
  Nauels, Y. Xia, V. Bex and P.M. Midgley, Eds., Cambridge University Press}}.

\bibitem[Jacquelin(2014)]{Jacquelin_2014}
Jacquelin, J., 2014: \textnormal{{R\'egressions et \'equations int\'egrales}
  (accessed July 2024 at
  https://scikit-guess.readthedocs.io/en/latest/\_downloads/4cd313a50f7e08ab81758ce0bd661bc3/Regressions-et-equations-integrales.pdf)}.

\bibitem[Kalman(1994)]{Kalman_1994}
Kalman, R.~E., 1994: \textnormal{Randomness reexamined}. {\em
  \textnormal{Modeling, Identification and Control}\/}, {\bf 15}, 141--151,
  doi:10.4173/mic.1994.3.3.

\bibitem[Kratzert et~al.(2019)Kratzert, Klotz, Shalev, Klambauer, Hochreiter,
  and Nearing]{Kratzert_etal_2019a}
Kratzert, F., D.~Klotz, G.~Shalev, G.~Klambauer, S.~Hochreiter, and G.~Nearing,
  2019: {Towards learning universal, regional, and local hydrological behaviors
  via machine learning applied to large-sample datasets}. {\em Hydrol. Earth
  System Sci.\/}, {\bf 23}, 5089--5110, doi:10.5194/hess--23--5089--2019.

\bibitem[Lehner et~al.(2008)Lehner, Verdin, and Jarvis]{Lehner_etal_2008}
Lehner, B., K.~Verdin, and A.~Jarvis, 2008: New global hydrography derived from
  spaceborne elevation data. {\em Eos Trans. AGU\/}, {\bf 89}, 93--94,
  doi:10.1029/2008EO100001.

\bibitem[Lehner et~al.(2011)Lehner, Liermann, Revenga, V{\"o}r{\"o}smarty,
  Fekete, Crouzet, D{\"o}ll, Endejan, Frenken, Magome, Nilsson, Robertson,
  R{\"o}del, Sindorf, and Wisser]{Lehner_etal_2011}
Lehner, B., C.~R. Liermann, C.~Revenga, C.~V{\"o}r{\"o}smarty, B.~Fekete,
  P.~Crouzet, P.~D{\"o}ll, M.~Endejan, K.~Frenken, J.~Magome, C.~Nilsson,
  J.~Robertson, R.~R{\"o}del, N.~Sindorf, and D.~Wisser, 2011: High-resolution
  mapping of the world's reservoirs and dams for sustainable river-flow
  management. {\em Front. Ecol. Environ.\/}, {\bf 9}, 494--502,
  doi:10.1890/100125.

\bibitem[Lorenz(1985)]{Lorenz_1985}
Lorenz, E.~N., 1985: \textnormal{{The growth of errors in prediction,
  Proceedings of the International School of Physics {\it Enrico Fermi},
  M.~Ghil, R.~Benzi, and G.~Parisi, Eds., Course 88 on Turbulence and
  Predictability in Geophysical Fluid Dynamics and Climate Dynamics, North
  Holland: Amsterdam, 243--265.}}

\bibitem[Mahfouf(1991)]{Mahfouf_1991}
Mahfouf, J.-F., 1991: Analysis of soil moisture from near-surface parameters: A
  feasibility study. {\em J. Appl. Meteor.\/}, {\bf 30}, 1534--1547,
  doi:10.1175/1520--0450(1991)030$<$1534:AOSMFN$>$2.0.CO;2.

\bibitem[Maraun(2016)]{Maraun_2016}
Maraun, D., 2016: {Bias correcting climate change simulations - a critical
  review}. {\em Curr. Clim. Change Rep.\/}, {\bf 2}, 211--220,
  doi:10.1007/s40641--016--0050--x.

\bibitem[Mauritsen et~al.(2019)Mauritsen, Bader, Becker, Behrens, Bittner,
  Brokopf, Brovkin, Claussen, Crueger, Esch, Fast, Fiedler, Fl{\"a}schner,
  Gayler, Giorgetta, Goll, Haak, Hagemann, Hedemann, Hohenegger, Ilyina, Jahns,
  {Jimen{\'e}z-de-la-Cuesta}, Jungclaus, Kleinen, Kloster, Kracher, Kinne,
  Kleberg, Lasslop, Kornblueh, Marotzke, Matei, Meraner, Mikolajewicz, Modali,
  M{\"o}bis, Müller, Nabel, Nam, Notz, Nyawira, Paulsen, Peters, Pincus,
  Pohlmann, Pongratz, Popp, Raddatz, Rast, Redler, Reick, Rohrschneider,
  Schemann, Schmidt, Schnur, Schulzweida, Six, Stein, Stemmler, Stevens, von
  Storch, Tian, Voigt, Vrese, Wieners, Wilkenskjeld, and Alexander
  Winkler~and]{Mauritsen_etal_2019}
Mauritsen, T., J.~Bader, T.~Becker, J.~Behrens, M.~Bittner, R.~Brokopf,
  V.~Brovkin, M.~Claussen, T.~Crueger, M.~Esch, I.~Fast, S.~Fiedler,
  D.~Fl{\"a}schner, V.~Gayler, M.~Giorgetta, D.~S. Goll, H.~Haak, S.~Hagemann,
  C.~Hedemann, C.~Hohenegger, T.~Ilyina, T.~Jahns,
  D.~{Jimen{\'e}z-de-la-Cuesta}, J.~Jungclaus, T.~Kleinen, S.~Kloster,
  D.~Kracher, S.~Kinne, D.~Kleberg, G.~Lasslop, L.~Kornblueh, J.~Marotzke,
  D.~Matei, K.~Meraner, U.~Mikolajewicz, K.~Modali, B.~M{\"o}bis, W.~A.
  Müller, J.~E. M.~S. Nabel, C.~C.~W. Nam, D.~Notz, S.-S. Nyawira, H.~Paulsen,
  K.~Peters, R.~Pincus, H.~Pohlmann, J.~Pongratz, M.~Popp, T.~J. Raddatz,
  S.~Rast, R.~Redler, C.~H. Reick, T.~Rohrschneider, V.~Schemann, H.~Schmidt,
  R.~Schnur, U.~Schulzweida, K.~D. Six, L.~Stein, I.~Stemmler, B.~Stevens,
  J.-S. von Storch, F.~Tian, A.~Voigt, P.~Vrese, K.-H. Wieners,
  S.~Wilkenskjeld, and E.~R. Alexander Winkler~and, 2019: {Developments in the
  MPI‐M Earth System Model version 1.2 (MPI‐ESM1.2) and its response to
  increasing CO$_2$}. {\em J. Adv. Model. Earth Sys.\/}, {\bf 11}, 998--1038,
  doi:10.1029/2018MS001400.

\bibitem[Meehl et~al.(2012)Meehl, Washington, Arblaster, Hu, Teng, Tebaldi,
  Sanderson, Lamarque, Conley, Strand, and {White III}]{Meehl_etal_2012}
Meehl, G.~A., W.~M. Washington, J.~M. Arblaster, A.~Hu, H.~Teng, C.~Tebaldi,
  B.~N. Sanderson, J.-F. Lamarque, A.~Conley, W.~G. Strand, and J.~B. {White
  III}, 2012: {Climate system response to external forcings and climate change
  projections in CCSM4}. {\em J. Climate\/}, {\bf 25}, 3661--3683,
  doi:10.1175/JCLI--D--11--00240.1.

\bibitem[Mendoza et~al.(2015)Mendoza, Clark, Barlage, Rajagopalan, Samaniego,
  Abramowitz, and Gupta]{Mendoza_etal_2015}
Mendoza, P.~A., M.~P. Clark, M.~Barlage, B.~Rajagopalan, L.~Samaniego,
  G.~Abramowitz, and H.~Gupta, 2015: {Are we unnecessarily constraining the
  agility of complex process-based models?} {\em Water Resour. Res.\/}, {\bf
  51}, 716--728, doi:10.1002/2014WR015820.

\bibitem[Messager et~al.(2016)Messager, Lehner, Grill, Nedeva, and
  Schmitt]{Messager_etal_2016}
Messager, M.~L., B.~Lehner, G.~Grill, I.~Nedeva, and O.~Schmitt, 2016:
  {Estimating the volume and age of water stored in global lakes using a
  geo-statistical approach}. {\em Nat. Commun.\/}, {\bf 7}, 1--11,
  doi:10.1038/ncomms13603.

\bibitem[Moriasi et~al.(2015)Moriasi, Gitau, Pai, and
  Daggupati]{Moriasi_etal_2015}
Moriasi, D.~N., M.~W. Gitau, N.~Pai, and P.~Daggupati, 2015: {Hydrologic and
  water quality models: Performance measures and evaluation criteria}. {\em
  Trans. ASABE\/}, {\bf 58}, 1763--1785, doi:10.13031/trans.58.10715.

\bibitem[Morin and Champoux(2006)]{Morin_Champoux_2006}
Morin, J., and O.~Champoux, 2006: \textnormal{Integrated modelling of the
  physical processes and habitats of the St.~Lawrence River, Chapter~3 of {{\it
  Water Availability Issues for the St.~Lawrence River: An Environmental
  Synthesis}}, {A.~Talbot, Ed., Environment Canada}, 22--37}.

\bibitem[Nearing et~al.(2016)Nearing, Tian, Gupta, Clark, Harrison, and
  Weijs]{Nearing_etal_2016}
Nearing, G.~S., Y.~Tian, H.~V. Gupta, M.~P. Clark, K.~W. Harrison, and S.~V.
  Weijs, 2016: {A philosophical basis for hydrological uncertainty}. {\em
  Hydrol. Sci. J.\/}, {\bf 61}, 1666--1678, doi:10.1080/02626667.2016.1183009.

\bibitem[{O’Neill} et~al.(2017){O’Neill}, Kriegler, Ebi, {Kemp-Benedict},
  Riahi, Rothman, {van Ruijven}, {van Vuuren}, Birkmann, Kok, Levy, and
  Solecki]{ONeill_etal_2017}
{O’Neill}, B.~C., E.~Kriegler, K.~L. Ebi, E.~{Kemp-Benedict}, K.~Riahi, D.~S.
  Rothman, B.~J. {van Ruijven}, D.~P. {van Vuuren}, J.~Birkmann, K.~Kok,
  M.~Levy, and W.~Solecki, 2017: {The roads ahead: Narratives for shared
  socioeconomic pathways describing world futures in the 21st century}. {\em
  Global Environ. Change\/}, {\bf 42}, 169--180,
  doi:10.1016/j.gloenvcha.2015.01.004.

\bibitem[Parker(2017)]{Parker_2017b}
Parker, W.~S., 2017: {Computer simulation, measurement, and data assimilation}.
  {\em Brit. J. Phil. Sci.\/}, {\bf 68}, 273--304, doi:10.1093/bjps/axv037.

\bibitem[Pellerin and {Nzokou Tanekou}(2020)]{Pellerin_NzokouTanekou_2020}
Pellerin, J., and F.~{Nzokou Tanekou}, 2020: \textnormal{{Reference Hydrometric
  Basin Network update, Environment and Climate Change Canada, Gatineau,
  Qu{\'e}bec, 30 pp., available at
  https://collaboration.cmc.ec.gc.ca/cmc/hydrometrics/www/RHBN/RHBN\_EN.pdf
  (accessed October 10, 2023).}}

\bibitem[{Rafieei Nasab} et~al.(2020){Rafieei Nasab}, Karsten, Dugger,
  FitzGerald, Cabell, Gochis, Yates, Sampson, McCreight, Read, Zhang, and
  McAllister]{RafieeiNasab_etal_2020}
{Rafieei Nasab}, A., L.~Karsten, A.~Dugger, K.~FitzGerald, R.~Cabell,
  D.~Gochis, D.~Yates, K.~Sampson, J.~McCreight, L.~Read, Y.~Zhang, and
  M.~McAllister, 2020: \textnormal{{Overview of National Water Model
  calibration: General strategy and optimization, NCAR WRF-Hydro community
  training material (November 6), 28~slides, available at
  https://ral.ucar.edu/sites/default/files/public/projects/wrf-hydro/training-materials/calibrationnov2020-arezoo.pdf
  (accessed October 2, 2023).}}

\bibitem[Rumelhart et~al.(1986)Rumelhart, Hinton, and
  Williams]{Rumelhart_etal_1986}
Rumelhart, D.~E., G.~E. Hinton, and R.~J. Williams, 1986: Learning
  representations by back-propagating errors. {\em Nature\/}, {\bf 323},
  533--536, doi:10.1038/323533a0.

\bibitem[Schmidt et~al.(2017)Schmidt, Bader, Donner, Elsaesser, Golaz, Hannay,
  Molod, Neale, and Saha]{Schmidt_etal_2017}
Schmidt, G.~A., D.~Bader, L.~J. Donner, G.~S. Elsaesser, J.-C. Golaz,
  C.~Hannay, A.~Molod, R.~B. Neale, and S.~Saha, 2017: {Practice and philosophy
  of climate model tuning across six US modeling centers}. {\em Geosci. Model
  Dev.\/}, {\bf 10}, 3207--3223, doi:10.5194/gmd--10--3207--2017.

\bibitem[Smith(2006)]{Smith_2006}
Smith, L.~A., 2006: \textnormal{Predictability past, predictability present,
  {{\it Predictability of Weather and Climate}}, {T.~Palmer and R.~Hagedorn,
  Eds., Cambridge University Press}, 217--250, doi:10.2307/j.ctv1pnc1q9.10}.

\bibitem[Stadnyk et~al.(2020)Stadnyk, MacDonald, Tefs, D{\'e}ry, Koenig,
  Gustafsson, Isberg, and Arheimer]{Stadnyk_etal_2020}
Stadnyk, T.~A., M.~K. MacDonald, A.~Tefs, S.~J. D{\'e}ry, K.~Koenig,
  D.~Gustafsson, K.~Isberg, and B.~Arheimer, 2020: {Hydrological modeling of
  freshwater discharge into Hudson Bay using HYPE}. {\em Elem. Sci. Anth.\/},
  {\bf 8}, 1--18, doi:10.1525/elementa.439.

\bibitem[Stadnyk et~al.(2021)Stadnyk, Tefs, Broesky, D{\'e}ry, Myers, Ridenour,
  Koenig, Vonderbank, and Gustafsson]{Stadnyk_etal_2021}
Stadnyk, T.~A., A.~Tefs, M.~Broesky, S.~J. D{\'e}ry, P.~G. Myers, N.~A.
  Ridenour, K.~Koenig, L.~Vonderbank, and D.~Gustafsson, 2021: {Changing
  freshwater contributions to the Arctic: A 90-year trend analysis
  (1981–2070)}. {\em Elem. Sci. Anth.\/}, {\bf 9}, 1--26,
  doi:10.1525/elementa.2020.00098.

\bibitem[Stiles et~al.(2014)Stiles, Danielson, Poulsen, Brennan,
  Hristova-Veleva, Shen, and Fore]{Stiles_etal_2014}
Stiles, B.~W., R.~E. Danielson, W.~L. Poulsen, M.~J. Brennan, S.~M.
  Hristova-Veleva, T.-P. Shen, and A.~G. Fore, 2014: {Optimized tropical
  cyclone winds from QuikSCAT: A neural network approach}. {\em IEEE Trans.
  Geosci. Remote Sens.\/}, {\bf 52}, 7418--7434, doi:10.1109/TGRS.2014.2312333.

\bibitem[Taylor et~al.(2012)Taylor, Stouffer, and Meehl]{Taylor_etal_2012}
Taylor, K.~E., R.~J. Stouffer, and G.~A. Meehl, 2012: {An overview of CMIP5 and
  the experiment design}. {\em Bull. Amer. Meteor. Soc.\/}, {\bf 93}, 485--498,
  doi:10.1175/JTECH--D--12--00008.1.

\bibitem[Wang et~al.(2019)Wang, Wang, Rao, Orr, Yan, and
  Kotamarthi]{Wang_etal_2019}
Wang, J., C.~Wang, V.~Rao, A.~Orr, E.~Yan, and R.~Kotamarthi, 2019: {A parallel
  workflow implementation for PEST version 13.6 in high-performance computing
  for WRF-Hydro version 5.0: A case study over the midwestern United States}.
  {\em Geosci. Model Dev.\/}, {\bf 12}, 3523--3539,
  doi:10.5194/gmd--12--3523--2019.

\bibitem[{Water Survey of Canada}(2023)]{Hydat_2023}
{Water Survey of Canada}, 2023: \textnormal{{The National Water Data Archive
  (HyDAT, updated on October~24, 2022), Environment and Climate Change Canada,
  available at https://collaboration.cmc.ec.gc.ca/cmc/hydrometrics/www
  (accessed November 9, 2022).}}

\bibitem[Yamazaki et~al.(2019)Yamazaki, Ikeshima, Sosa, Bates, Allen, and
  Pavelsky]{Yamazaki_etal_2019}
Yamazaki, D., D.~Ikeshima, J.~Sosa, P.~D. Bates, G.~H. Allen, and T.~M.
  Pavelsky, 2019: {MERIT Hydro: A high‐resolution global hydrography map
  based on latest topography dataset}. {\em Water Resour. Res.\/}, {\bf 55},
  5053--5073, doi:10.1029/2019WR024873.

\bibitem[Zhang et~al.(2019{\natexlab{a}})Zhang, Perrie, and
  Long]{Zhang_etal_2019b}
Zhang, M., W.~Perrie, and Z.~Long, 2019{\natexlab{a}}: {Sensitivity study of
  North Atlantic summer cyclone activity in dynamical downscaled simulations}.
  {\em J. Geophys. Res. Atmos.\/}, {\bf 124}, 7599--7616,
  doi:10.1029/2018JD029766.

\bibitem[Zhang et~al.(2019{\natexlab{b}})Zhang, Flato, Kirchmeier-Young,
  Vincent, Wan, Wang, Rong, Fyfe, Li, and Kharin]{Bush_Lemmen_Zhang_etal_2019}
Zhang, X., G.~Flato, M.~Kirchmeier-Young, L.~Vincent, H.~Wan, X.~Wang, R.~Rong,
  J.~Fyfe, G.~Li, and V.~V. Kharin, 2019{\natexlab{b}}: \textnormal{{Changes in
  temperature and precipitation across Canada, Chapter~4 of {\it Canada's
  Changing Climate Report}, E. Bush and D. S. Lemmen, Eds., Government of
  Canada, Ottawa, Ontario, p.~112--193, (accessed April 2024 at
  https://changingclimate.ca/CCCR2019)}}.

\end{thebibliography}

\end{document}